\documentstyle[11pt,aaspp4,here]{article} 

\received{03 February 1999}
\accepted{07 April 1999}
 
\slugcomment{To appear in July 1999 issue of {\it The Astronomical Journal}} 
  
\lefthead{Ratnatunga, Griffiths, Ostrander}
\righthead{HST Medium Deep Survey database}

\begin{document}

\def\degpt {{$\buildrel{\circ} \over .$}}
\def\DPB{\rm D+B\ }
\def\BTD{{\rm B/T\ }}
\def\HLF{{\rm HLF }\ }
\def\HLPX{\xi\ }
\def\SNRIL{\Xi\ }
\def\etal{{et al.}}
\def\PAV3 {${\rm PA}_{\rm V_3}\ $}
\def\and{\rm and\ }
\def\ltorder{\mathrel{\raise.3ex\hbox{$<$}\mkern-14mu
             \lower0.6ex\hbox{$\sim$}}}
\def\gtorder{\mathrel{\raise.3ex\hbox{$>$}\mkern-14mu
             \lower0.6ex\hbox{$\sim$}}}

\title {Disk and Bulge Morphology of WFPC2 galaxies: \\
The HST Medium Deep Survey database}

\author{ Kavan U. Ratnatunga, Richard E. Griffiths, Eric J. Ostrander}

\affil{Physics Dept., Carnegie Mellon University, \\
Pittsburgh, PA 15213 \\
 kavan, griffith \& ejo@astro.phys.cmu.edu}
\begin{abstract}

Quantitative morphological and structural parameters are estimated for
galaxies detected in HST observations of WFPC2 survey fields.
A modeling approach based on maximum likelihood has been developed for
two-dimensional decomposition of faint under-sampled galaxy images
into components of disk and bulge morphology. Decomposition can be
achieved for images down to F814W (I) $\approx 23.0$, F606W (V)
$\approx 23.8$ and F450W (B) $\approx 23.3$ magnitudes in WFPC2
exposures of one hour. We discuss details of the fitting procedure,
and present the observed distributions of magnitude, color, effective
half-light radius, disk and bulge axis ratios, bulge/(disk+bulge) flux
ratio, bulge/disk half-light radius ratio and surface brightness. We
also discuss the various selection limits on the measured parameters.
The Medium Deep Survey catalogs and images of random pure parallel fields
and other similar archival primary WFPC2 fields have been made available
via the Internet with a searchable browser interface to the database.
\footnote {at http://archive.stsci.edu/mds/}

\end{abstract}


\section{Introduction } 

WFPC2 pure parallel images from the HST Medium Deep Survey key project
(\cite{1994ApJ...437...67G,1994ApJ...435L..19G} hereafter MDS ) cover
a very wide range of signal-to-noise. For the few brightest galaxies
observed, detailed structures such as spiral arms and bright regions
of star-formation are well exposed and the morphology can be easily
classified by eye and measured by traditional interactive
one-dimensional profile fitting procedures. At these brighter
magnitudes the two-dimensional light distributions of galaxies are not
well fitted by simple parameterized models which are necessarily crude
fits to the broad continuum using smooth image profiles. However, as
the images get fainter and smaller (undersampled), the morphology is
less apparent and requires a model-based two-dimensional image
analysis to derive quantitative estimates. For the extreme faint and
small objects there is very little morphological information in the
observations. The MDS procedure described in this paper has been
optimized for the intermediate (medium deep) galaxies, in the rough
magnitude range between V $\approx 21$ to 24 mag., as imaged in exposures of
about one hour. This has yielded a significantly large catalog of
quantitative morphological and structural parameter estimates. This
magnitude range is now accessible for spectroscopic determination of
redshifts via the new generation of 8-10 meter class ground based
telescopes.

Decomposition of the images into disk and bulge has been a difficult
task even at bright magnitudes with well sampled images
(\cite{1977ApJ...217..406K,1981ApJS...46..177B,1984ApJS...56..105K,1985ApJS...59..115K}).
Interactive procedures (\cite{1991PASP..103..396Y}) are also
impractical for a large survey and in any case they do not generate an
uniform catalog suitable for statistical analysis. The image analysis
adopted is similar to that in stellar photometry programs like DAOphot
(\cite{1987PASP...99..191S}). But unlike stellar photometry where the
image can be characterized by the centroid, magnitude and the Point
Spread Function (PSF), there is no simple model which will
intrinsically fit all of the galaxy images. We adopt axisymmetric
scale-free models which have been shown to fit the image continuum of
normal galaxies (\cite{1959Hbphy..53..275V,1970ApJ...160..811F}). The
procedure will average over any bright regions, as typically occurs in
the data themselves at fainter magnitudes where the objects are
smaller and less resolved. The residuals to these simple galaxy model
fits at brighter magnitudes are the subject of a separate study
(\cite{1997ApJ...476..510N}). To limit the complexity of the analysis,
we assume that an image pixel is associated with a single object or
background sky as is typical of the MDS WFPC2 images. We do not deal
with the problems of crowding or image overlap, which are the major
issues in programs for stellar photometry.

The number and choice of parameters fitted to an extended image is
clearly important. Fitting too few parameters to a well exposed image
could significantly bias the estimates of the parameters fitted, by
the implicit choice of the parameters that are not fitted. However,
fitting too many parameters to faint and/or compact unresolved images
could cause the fit to converge to a false local minimum of a
likelihood function which is very noisy in that multidimensional
space. For practical reasons, and to ensure statistical uniformity of
the resulting catalog, we require an automated procedure which will
select and fit those (necessary and sufficient) parameters which are
constrained by each particular image. We have developed
two-dimensional ``maximum likelihood'' image analysis software that
attempts to automatically optimize the model and the number of
parameters fitted to each image. We apply the Ockham's razor: {\it
non sunt multiplicanda entia praeter necessitatem; i.e., } entities
are not to be multiplied beyond necessity (Ockham 1285-1348). The
model varies from a simultaneous decomposition of disk and bulge
components of galaxy images ( hereafter \DPB models) at the bright end
to circularly symmetric sources at the faint end. However, this choice
of parameters creates selection effects which depend on the
signal-to-noise of the image and needs to be included explicitly in
any statistical analysis of the MDS database.

The success of the procedure depends on the ability to efficiently
generate smooth subpixelated galaxy images which can be convolved with
an adopted Point Spread Function (PSF), such that precise derivatives
can be evaluated with respect to all the parameters which need to be
estimated. We will outline the procedure here but will avoid giving
all the details of the numerical algorithm, since they are probably
not of interest to the general reader. The algorithm is documented by
comments in the software and the interested reader should contact the
first author. A brief outline of the MDS pipeline is given in the
Appendix.

This paper is also the primary reference to the Medium Deep Survey
database which has been made available on the MDS website in the HST
archive \footnote{at http://archive.stsci.edu/mds/} and also mirrored
at the Canadian Astronomy Data Center (CADC)
\footnote{at http://cadcwww.dao.nrc.ca/mds/}. We avoid duplicating
extensive tables since those can only be a snapshot of the present MDS
database and we wish to ensure that users will always refer to the
latest version which will be maintained on the Internet. The MDS
website has a cgi-interface written in f77 which allows the database
to be searched using coordinates or galaxy parameters, or looked at
interactively by clicking on objects on an image-map of each
stack. Direct access is also provided to the MDS database which is on
CDROMs in a `jukebox'.

The database contains WFPC2 Pure Parallel observations taken for the
Medium Deep Survey (MDS - HST GO program ids 5369, 5370, 5371, 5372,
5971, 6251, 6802, 7203) and for the GTO observers (HST program ids
5091, 5092, 5201, 6252, 6254, 6609, 6610, 7202 ) as well as HST
archival observations of randomly selected WFPC2 fields like that of
the Groth-Westphal strip (HST GTO program ids 5090 5109 - hereafter
GWS ) and the Hubble Deep Field (HST DD program id 6337 - hereafter
HDF ), and selected galaxy cluster fields (HST archival program id
7536) and will continue to be expanded as more fields are processed
(HST archival program id 8384).

\section{The Observations}

The HST MDS and GTO pure parallel observations were taken with the
WFPC2 after January 1994, following the SM93 repair mission, and
continued for four years until January 1998. Before the SM97 second
servicing mission in February 1997, the instruments used for the
associated HST primary observations were the FGS, FOC and FOS; after
this mission, the primary instruments were FGS, STIS and NICMOS.


\begin{figure}
\plotfiddle{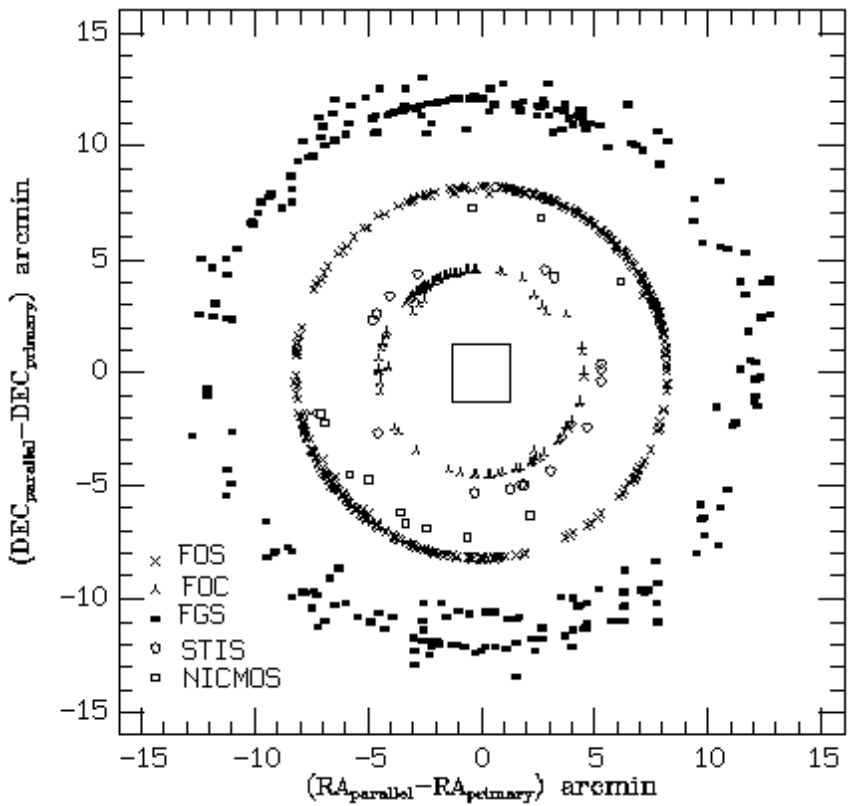}{6.0in}{-90.}{75.}{75.}{-288}{432}
\medskip
\caption{
Location of pure parallel fields relative to targets in primary
observations. Instrument of HST primary as indicated by symbols. 
\label{fig1}}
\end{figure}

We illustrate in Figure~\ref{fig1} the difference in pointing between
the parallel and primary observations for all pure parallel fields in
the MDS database, using different symbols for each primary
instrument. The WFPC2 field is on average 4.5, 8.2, 12.1, 7.1, and 5.3
arc min away from the FOC, FOS, FGS, STIS, and NICMOS primary target
respectively.
 
\begin{figure}
\plotfiddle{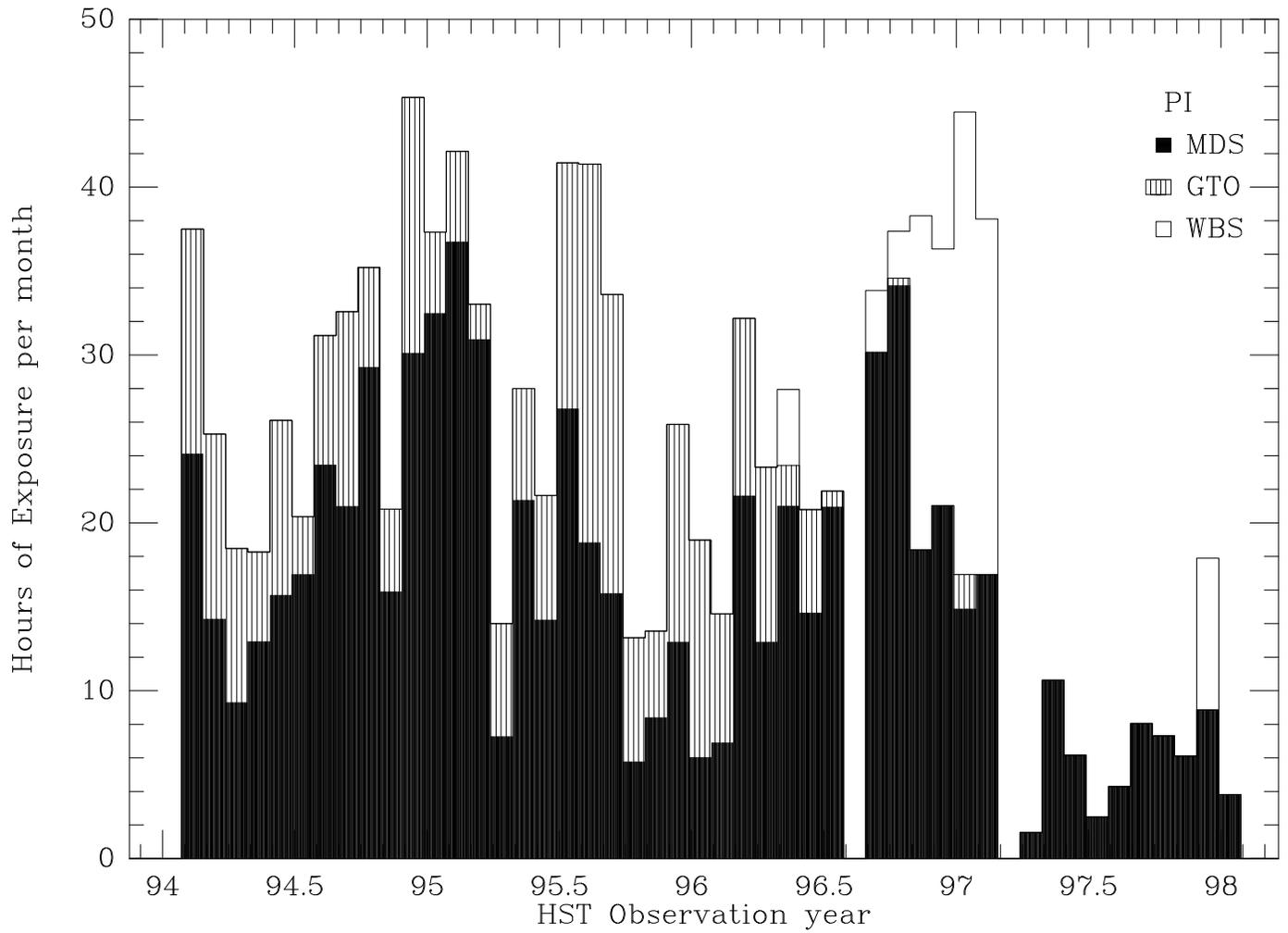}{6.0in}{-90.}{75.}{75.}{-288}{432}
\medskip
\caption{
Hours of pure parallel WFPC2 observations
over the 4 years of the Medium Deep Survey. Program as indicated by shading.
\label{fig2}}
\end{figure}

About 25 hours of pure parallel exposure was obtained each month,
giving a steady flow of observations. The database was supplemented
using the archival data from primary observations which satisfied the
survey criteria.

The observation history is illustrated in Figure~\ref{fig2} as an
indication of the quantity of HST data that was available for the
survey. There was a significant drop in the number of WFPC2 parallels
after SM97 as a result of the dithered observing strategy of NICMOS and
STIS primary observations. The pure parallel GTO data was available to
REG as a WFPC2 Investigation Definition Team member and Windhorst's
Blue Survey data (WBS) was available from the HST archive 3 months
after observation.

\begin{figure}
\plotfiddle{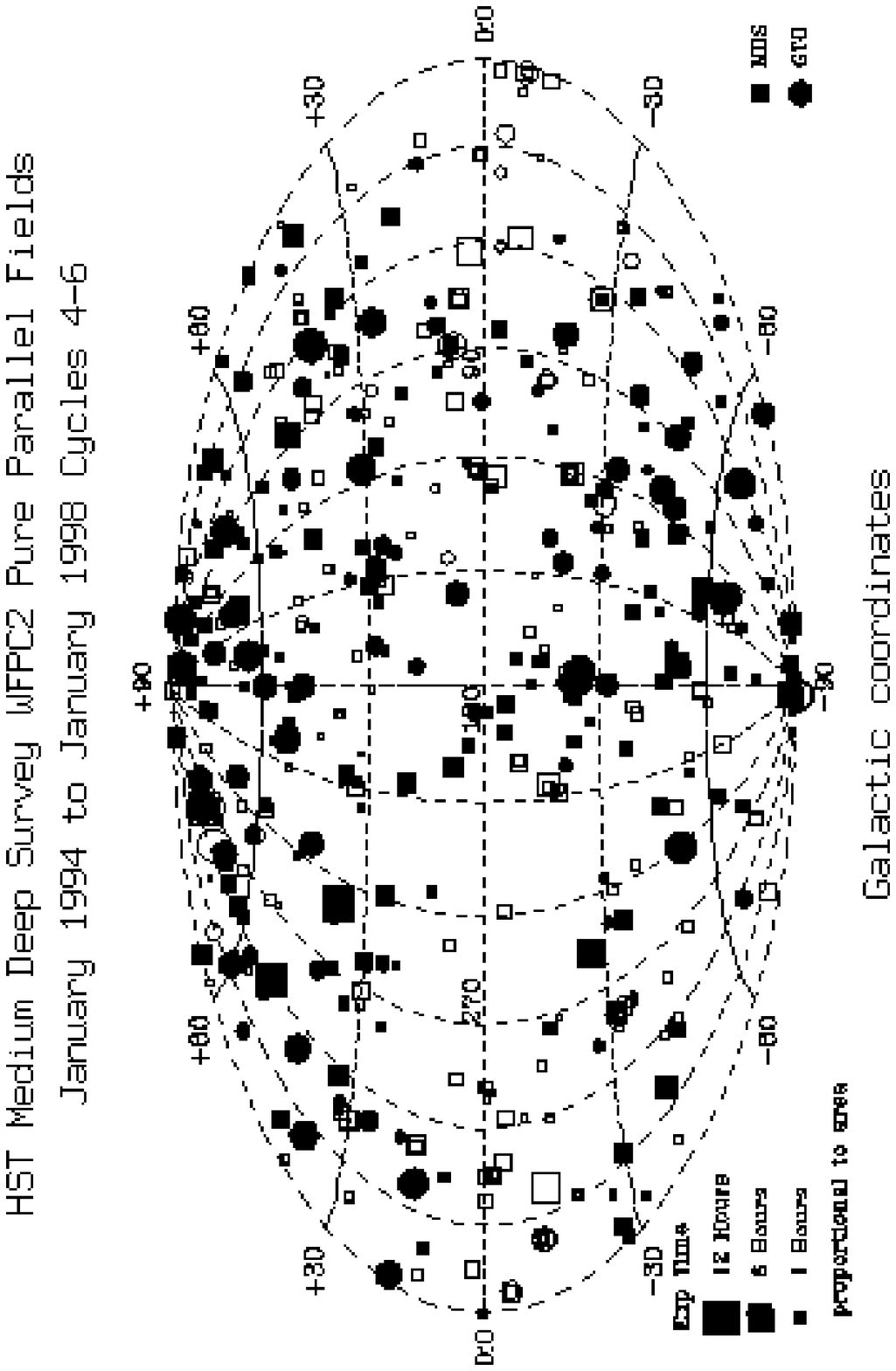}{6.0in}{-90.}{75.}{75.}{-288}{432}
\medskip
\caption{
Distribution of the MDS pure parallel WFPC2 fields over the sky
in Galactic Coordinates. Symbol size proportional to total length of exposure
and are filled for fields processed
for MDS.
\label{fig3}}
\end{figure}

The HST MDS (\cite{1994ApJ...435L..19G}), with over 400 random WFPC2
fields distributed over the full Sky, the GWS
(\cite{1994AAS...185.5309G}) with 28 contiguous WFPC2 fields and the
Hubble Deep Field (HDF \cite{1996AJ....112.1335W}) are datasets which
give three very complementary samples of field galaxies at faint
magnitude. The HDF gives depth in a single WFPC2 field, the GWS gives
a larger area uniformly observed, and the MDS samples the whole sky as
illustrated in Figure~\ref{fig3}. All three sets have been analyzed
uniformly through the MDS pipeline analysis software system.

\slugcomment{ } 
\begin{center}
\begin{table}[H]
\title{MDS Field Priority}
\medskip
\begin{tabular}{ c l }
\tableline
Priority & Description \\
\tableline
\tableline
\\
1 & 3 or more images in each of 2 or more Filters \\
2 & 2 or more images in each of 2 or more Filters \\
3 & 3 or more images in 1 Filter \\
4 & 2 images in one and 1 image in other Filter \\
5 & 1 image each in 2 Filters \\
6 & 2 images in 1 Filter \\
7 & 1 image only \\
8 & Bad exposure \\
9 & Failed observation \\
\tableline
\end{tabular} 
\caption{Assignment of Field Priority in the MDS pipeline\label{tbl-1}}
\end{table} 
\end{center} 

MDS and GTO observations were primarily done with the F814W and F606W
broadband filters. When more than 3 exposures could be taken with each
of these filters, then F450W observations were taken in addition to
those in the first two. In order to achieve a similar signal-to-noise
ratio in the images taken in all three filters, the exposure times in
F814W and F606W were requested to be about equal while in F450W the
requested exposure was about twice as long. However, all WFPC2
observations in the MDS were taken in ``non-interference'' pure
parallel mode (\cite{1994ApJ...437...67G}), with the result that
exposure times were of varied duration, with a variable number of
exposures in the stack used for cosmic ray removal.

\begin{figure}
\plotfiddle{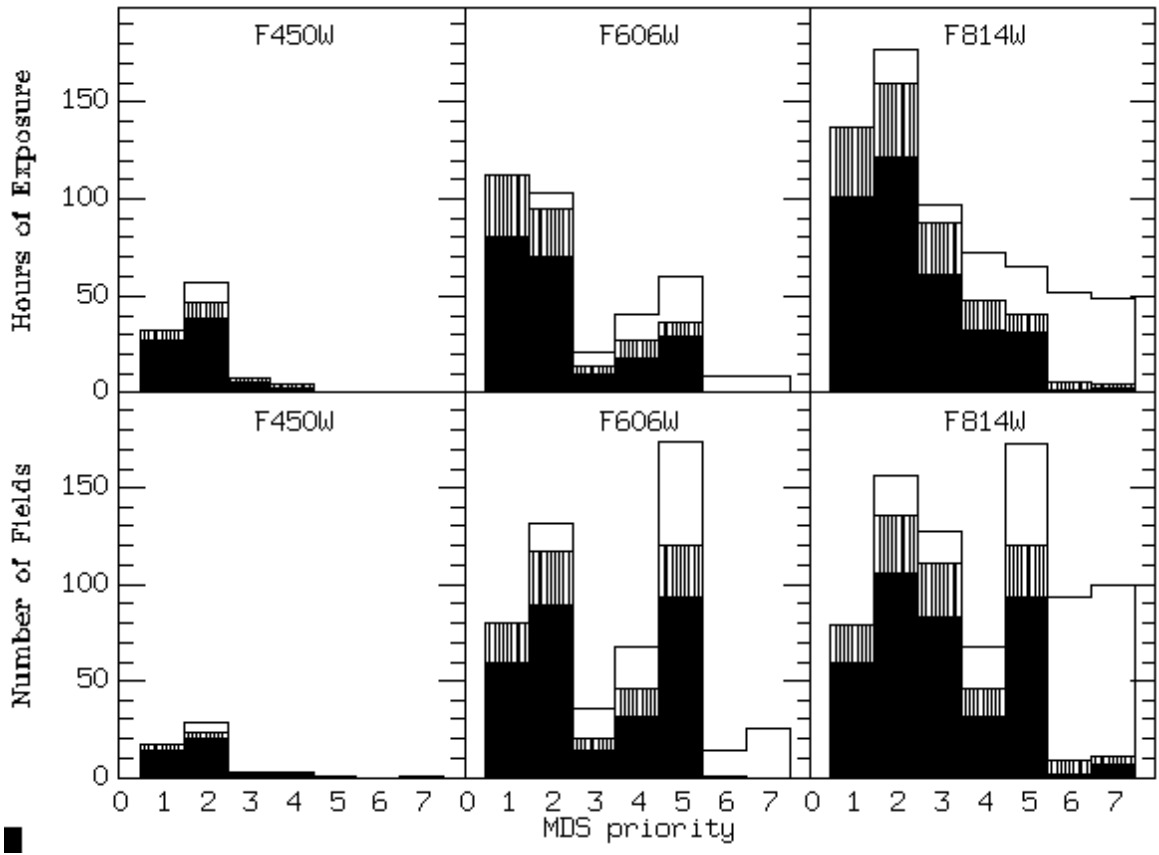}{6.0in}{-90.}{75.}{75.}{-288}{432}
\medskip
\caption{
The total hours of exposure and number of fields at each priority (see
Table~\ref{tbl-1}) in each of the 3 primary WFPC2 Filters. The filled
region of histogram are of fields processed through MDS pipeline. The
partial shaded region is of non-survey fields.
\label{fig4}}
\end{figure}

Each field was given a priority based on the number of exposures
available, as listed in Table~\ref{tbl-1}. The total hours of exposure
and the number of fields at each priority in each of the 3 selected
WFPC2 Filters is illustrated in Figure~\ref{fig4}. Practically all of
the higher priority data has been processed through the MDS pipeline
and made available in the MDS website. The single exposure, single filter
fields were given lowest priority because of the inability to remove
cosmic rays and the lack of any color information. After October 1995,
the MDS used only pure parallel opportunities in which a minimum of
two exposures with total exposure time longer than 20 min could be
taken in each of two filters, or one exposure with a total exposure
time longer than 30 min in each of two filters. Special data
processing code was developed to perform cosmic ray rejection using
exposures through different filters. This procedure, although better
than attempting to clean cosmic rays from a single exposure, is
performed at the cost of losing any objects of extreme color.

The pure parallel observations {\it per se} have therefore been the
biggest challenge in the task of building a database using a clean
statistical analysis. We will use the GWS for many of the illustrated
distributions, in order to avoid complicating the discussion with
effects due to changes in data quality.

\section{The signal-to-noise index $\SNRIL$. }

To characterize the ability of our method to extract quantitative
parameter estimates we define an information index based on the
signal-to-noise in the image. Since we are dealing with mostly
extended images, the definition of the index is different from the
signal-to-noise ratio generally used for point sources.

\begin{figure}
\plotfiddle{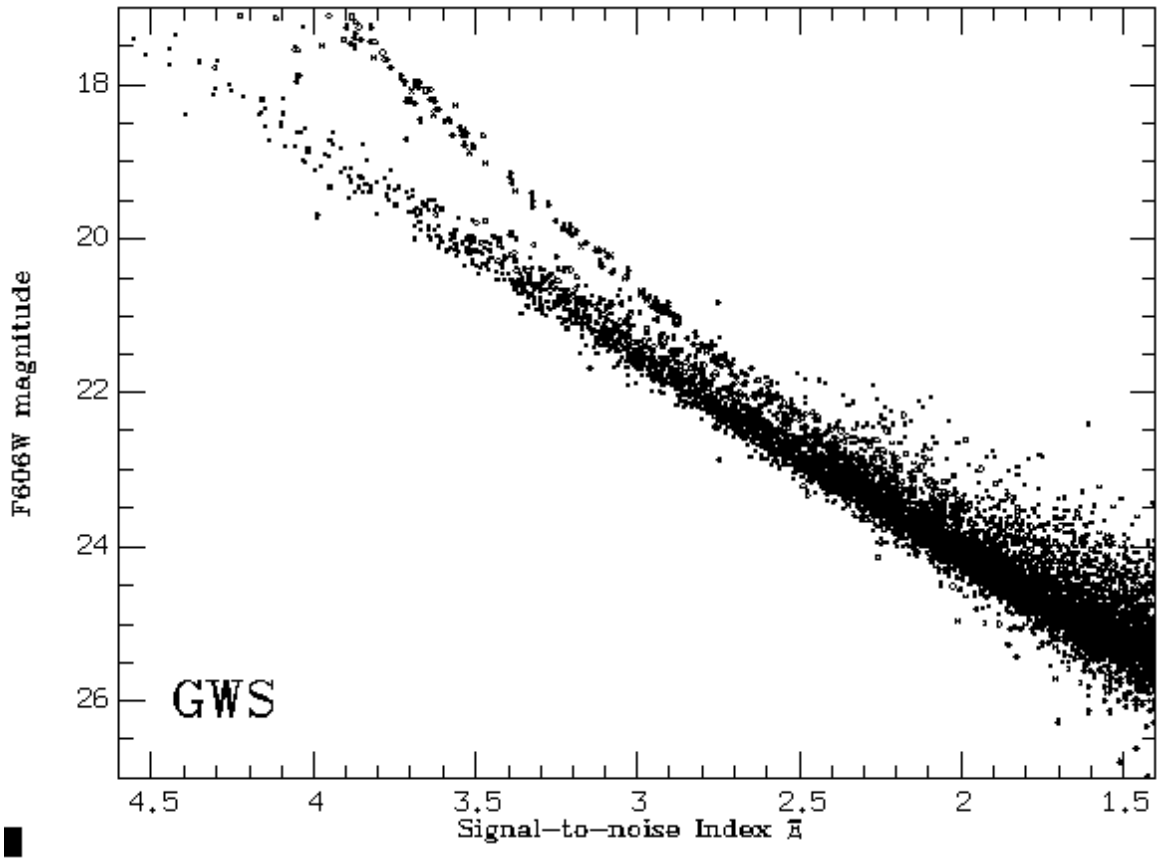}{6.0in}{-90.}{75.}{75.}{-288}{432}
\medskip
\caption{
The decimal logarithm of the integrated signal-to-noise ratios of
pixels within $1\sigma$ isophote ($\SNRIL$) of extended images in GWS is
linearly correlated with slope of 1 magnitudes per 0.4dex. Point-like
stars follow a different sequence at brighter magnitudes. Symbols represent
different models as indicated in Fig~22.
\label{fig5}}
\end{figure}

We first define a contour around an object by selecting the subset of
contiguous pixels which each contain a signal that is at least $1\sigma$
above the estimated local sky (see appendix for details). The
signal-to-noise ratio of each of these pixels is computed
individually. We define the signal-to-noise index $\SNRIL$ as the
decimal logarithm of the integral sum of these ratios, and we have
found this dimensionless quantity to be a good measure of the
information content of the image, and we have used it to define
thresholds within the image analysis procedure. For any particular
field, exposure time and WFPC2 filter, $\SNRIL$ is linearly correlated
with the magnitude of an extended image. Furthermore, it has the
expected slope of 1 magnitude per 0.4dex, as shown in
Figure~\ref{fig5} for GWS observations through F606W. Point-like
stars follow a different sequence at brighter magnitudes.

In most of the discussion on image analysis we will refer to $\SNRIL$
rather than magnitude since it is a measure of image quality which can
be used without reference to exposure time, sky background and filter
used.

The MDS detection limit is at $\SNRIL \approx 1.6$, but the sample
does contain some images with a smaller index, {\it viz.} those
objects detected in the image of a different filter of the same region
of sky. The completeness limit is at $\SNRIL \approx 1.8$ which is a
half magnitude brighter than the detection limit. The morphology
(disk-like or bulge-like) of galaxies can be determined for
$\SNRIL\gtorder 2.0$ and \DPB models can be done for $\SNRIL\gtorder
2.4$, which is 2 magnitudes brighter than the detection limit. To
avoid any contamination by image noise the detection limit is set at a
conservative level since that was already much fainter than the image
quality needed to estimate morphology.

\begin{figure}
\plotfiddle{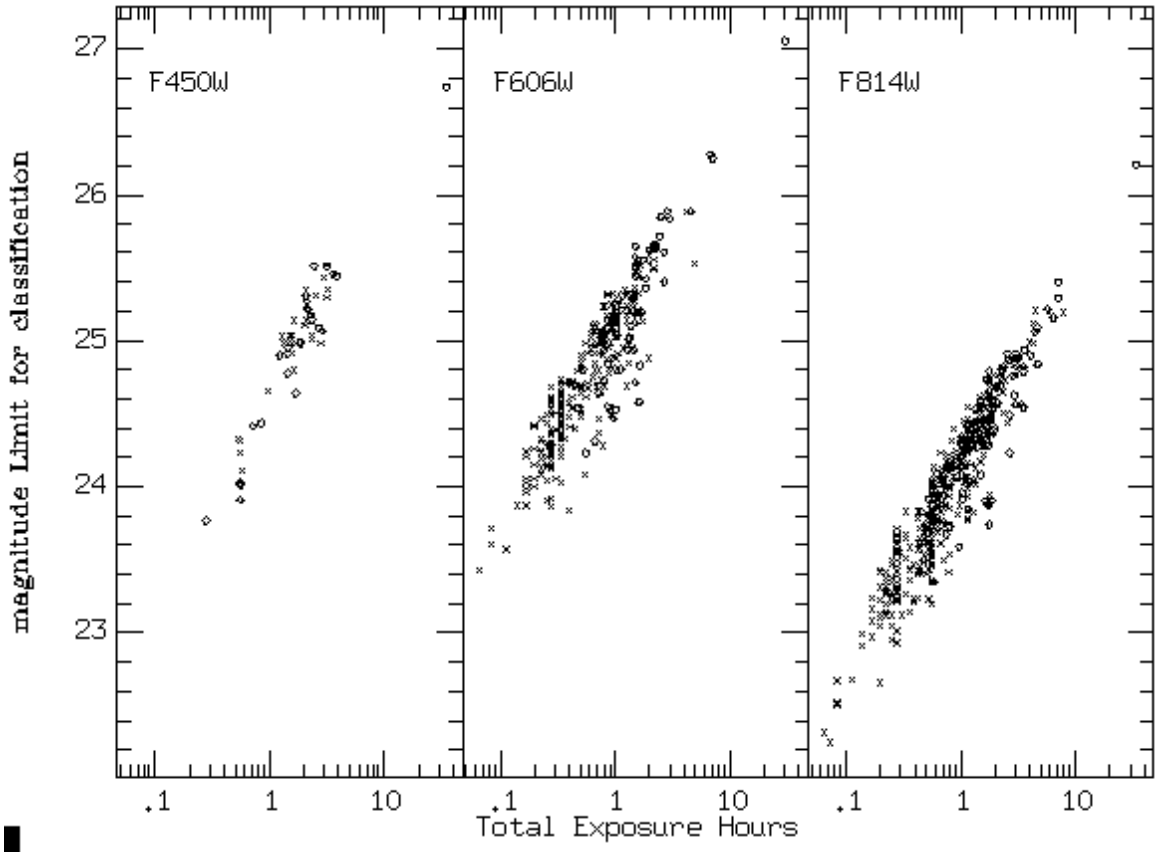}{6.0in}{-90.}{75.}{75.}{-288}{432}
\medskip
\caption{
Limiting magnitudes for morphology classification ($\SNRIL\approx
1.8$) as a function of total exposure time for all fields processed in
3 primary WFPC2 Filters. The best fields with 3 or more images in
stack are plotted as circles. The northern HDF field is the extreme
point on each graph.
\label{fig6}}
\end{figure}

In Figure~\ref{fig6} we illustrate the limiting magnitude
($\SNRIL \approx 1.8$) as a function of total exposure time for all the MDS
fields processed from WFPC2 pure parallel observations from HST Cycles
4 through 6, the GWS, the HDF, and archival cluster fields.

The GWS comprises 27 WFPC2 fields, each observed uniformly with 4
exposures in each of the I (F814W) and V (F606W) filters, with total
integration times of 4400 and 2800 seconds respectively and one deep WFPC2 
field with $\approx 25,000$ seconds in each filter. Our object
catalog for the GWS has 12,800 objects in the 27 WFPC2 fields. The
percentage of images with $\SNRIL \gtorder 4.0,\ 3.0,\ 2.0\ \and 1.5$ is
$0.3\% , 4.6\% , 30\% \and 67\%\ $ respectively. In these survey
images, $\SNRIL\approx 1.8$ corresponds to I=24.5 mag and to V=25.2 mag. From
a catalog of $\approx 10,800 $ galaxy images with $\SNRIL \gtorder 1.8$ in both
F814W and F606W, 11\% of the images are fitted with two-component \DPB
models, 7\% are classified as stars, 61\% are classified as either
disk-like or bulge-like, 20\% are classified as generic galaxies (of
uncertain disk or bulge nature) and less than 1\% remain unclassified.

\begin{figure}
\plotfiddle{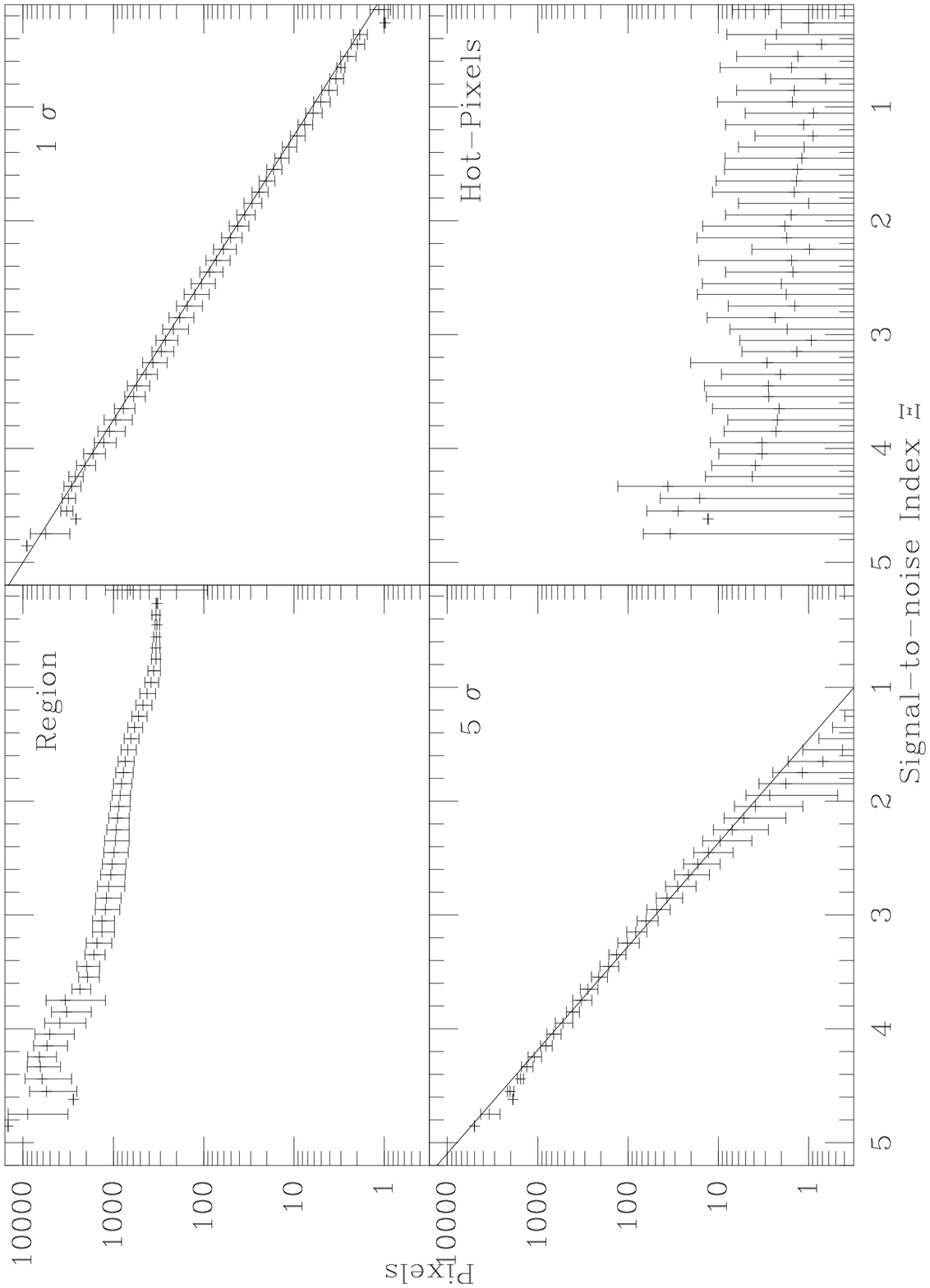}{6.0in}{-90.}{75.}{75.}{-288}{432}
\medskip
\caption{
Average number of pixels in selected region and above $5\sigma$ and
$1\sigma$ contours as a function of $\SNRIL$ of images in GWS.
We also show the number of
pixels rejected as suspected residual hot pixels.
\label{fig7}}
\end{figure}

As illustrated in Figure~\ref{fig7}, we find empirically that on
average there are $10^{0.8\ \SNRIL}$ pixels above the $1\sigma$
contour, and $0.025\ 10^{1.1\ \SNRIL}$ pixels above $5\sigma$. When
$\SNRIL\approx 2.0$ we thus have typically an image with 40 pixels
above $1\sigma$ and 4 pixels above $5\sigma$. At the detection limit
of our object-finding algorithm ($\SNRIL\approx 1.6$), we have
typically an image with 15 pixels above $1\sigma$ and 1 pixel above
$5\sigma$. Most images with $\SNRIL\ltorder 1.6$
are of regions corresponding to objects which were detected in another
filter and model fits on them are typically very poor.

\begin{figure}
\plotfiddle{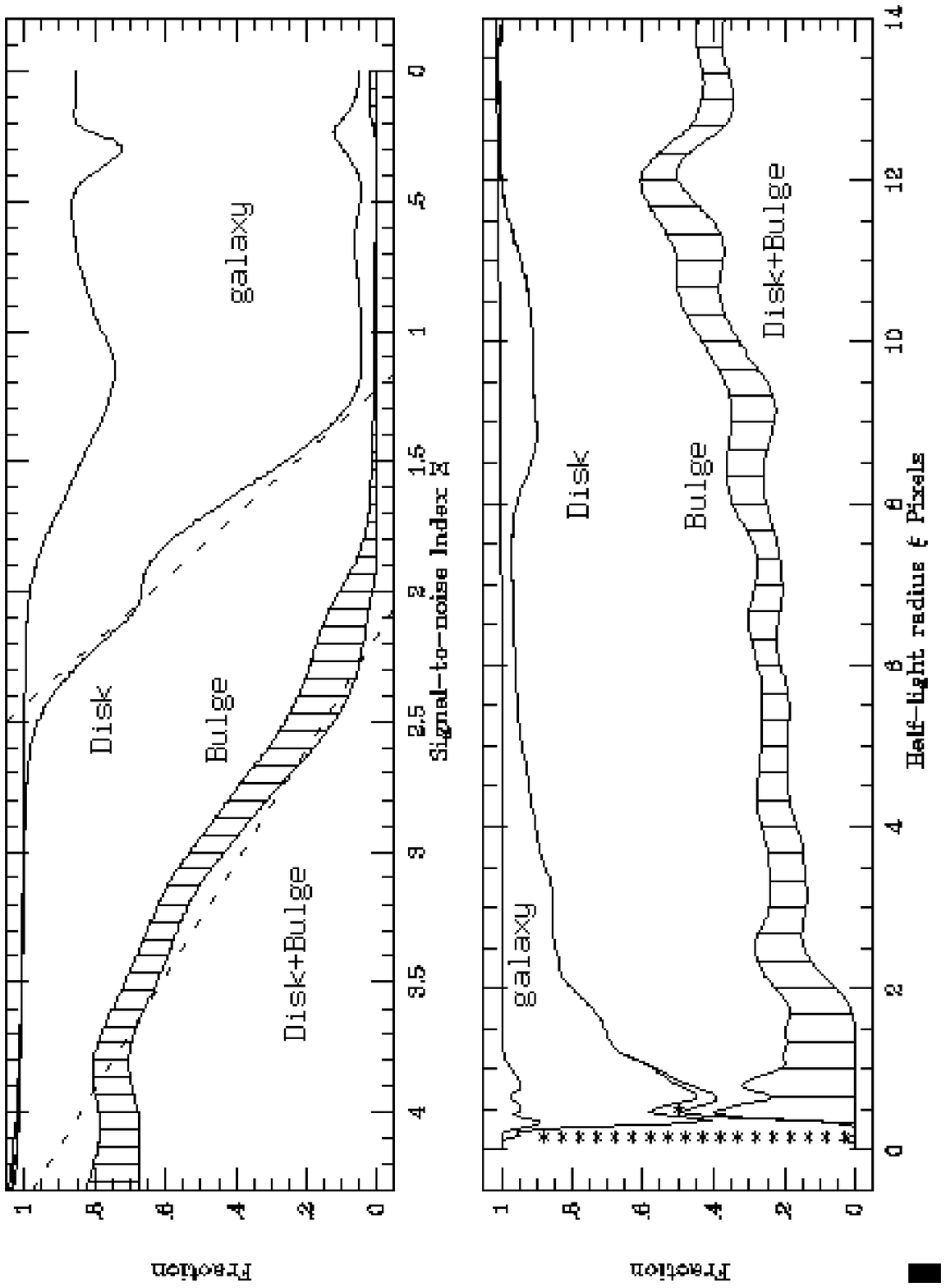}{6.0in}{-90.}{75.}{75.}{-288}{432}
\medskip
\caption{
Fraction of images in GWS fields classified as different
galaxy morphologies as a function of $\SNRIL$ (upper panel) and as a function
of half-light radius in pixels (lower panel) for images with $\SNRIL
>2.0$.
\label{fig8}}
\end{figure}

In Figure~\ref{fig8}a we show empirically that the fraction of images
for which we can use the likelihood ratio (see sec~7) to determine
whether the galaxy is more disk-like or bulge-like follows the
relation $\min(\max(0,(0.82\ \SNRIL-1)),1)$. Of these galaxies, the
fraction of images for which we can fit a significant two-component
model follows the relation $\min(\max(0,(0.46\ \SNRIL-1)),1)$. Both
these relations were derived by fitting a straight line to the slopes
in this figure. We can classify about 60\% of the galaxies with a
$\SNRIL \approx 2.0$, and all of them with $\SNRIL \gtorder
2.5$. Hardly any galaxy with $\SNRIL \ltorder 2.2$ has sufficient
signal to fit a two-component model, while 70\% of them can be fitted
at $\SNRIL \gtorder 3.5$, of which however there are only very few
examples in the MDS database. At $\SNRIL \approx 3.0$, about 40\% of
the galaxies are modeled as \DPB . The saturation of the fraction at
about 70\% is probably the intrinsic percentage of galaxies which have
a significant component in both disk and bulge.

In Figure~\ref{fig8}b, we show a plot similar to Figure~\ref{fig8}a,
as a function of the half-light radius in pixels for images with
$\SNRIL > 2.0$. For over 90\% of the images with a half-light radius
$\HLPX \gtorder 2$ pixels, the image can be classified statistically using
the likelihood ratio (see sec~7) to determine whether the galaxy is
more disk-like or bulge-like. We can fit a significant two-component
\DPB model to 20\%, 25\%, and 40\% of the galaxy images with 
$\HLPX \gtorder $ 2, 5 and 10 pixels respectively. None of the galaxies with 
$\HLPX \ltorder 2$ pixels had sufficient sampling to fit a two-component 
\DPB model fit.

The ability and success of fitting models to an observed galaxy depend
of course on both the integrated signal-to-noise index $\SNRIL$, as
well as the half-light radius of the galaxy in pixels $\HLPX$. These
two quantities are related to each other and to the morphology of the
image. Systematic (or evolutionary) changes in the mean size and
morphology as a function of apparent magnitude could slightly change
Figure~\ref{fig8} if it were to be drawn for WFPC2 fields at
significantly different limiting magnitudes. Figure~\ref{fig8} is
applicable to exposures of about 1-hour in F814W and F606W. The
difference in zero point magnitude for these filters is about the same
as the mean color of galaxies, and therefore we can expect similar
$\SNRIL$ for the typical galaxy. We have excluded in this figure the
galaxies imaged on the PC camera in order to keep the spatial
resolution constant. Each WFC pixel is $0\farcs1$.

\section{Maximum likelihood estimation}

Estimates of the centroid, magnitude, size, orientation and axis ratio
of the observed galaxy image are initially evaluated using simple
moments of the flux above the mean estimated sky, using those pixels
within the $1\sigma$ contour. We next select an elliptical region
around the object, ensuring that there are sufficient pixels to define
the mean sky background to $\approx$ 0.5\% accuracy (0.005 mag). Any
pixels within the elliptical region which are associated with some
other object and which are $1\sigma$ above the mean sky are cut out from
the region analyzed, together with any pixels which have been flagged
as ``bad'' in the calibration procedure (\cite{1994AJ....108.2362R}).

The procedure for the estimation of parameters via ``maximum
likelihood'' starts by initial estimates of the model parameters from
the observed moments of the image. For a given set of model
parameters, the software creates a model image of the object and
compares this image with the observations within the selected region
(including the error image). The ``likelihood function'' is defined as
the product of the probabilities for each model pixel value with
respect to the observed pixel value and its error distribution; this
function is evaluated as the integral sum of the logarithm of these
probabilities. The likelihood function is then maximized by using a
modified IMSL minimization routine (see \cite{1991AJ....101.1075R}).
The 2D-image analysis used an improved version of the software
developed for pre refurbishment WF/PC data
(\cite{1994STScI.RT2..333R}), the catalog of which was presented in
\cite{1995ApJ...453..599C}.

\section{Model fitting}

There are many types of empirical models that have been suggested over
the years to represent galaxy profiles. We have decided in particular
to use scale-free axisymmetric models with an exponential power-law
profile which have been shown to fit the broad continuum of normal
galaxies (\cite{1959Hbphy..53..275V,1970ApJ...160..811F}). This choice
has many numerical advantages which are desirable in leading towards
the development of a practical maximum likelihood fitting
algorithm. Elliptical galaxies are assumed to have a $e^{-r^{1/4}}$
(bulge-like) profile, and disk galaxies a $e^{-r}$ (disk-like)
profile. Each profile is characterized by a major axis half-light
radius and axis ratio. Some well exposed images need to be modeled as
the sum of two elliptical components.

For about 4\% of the galaxy images with no central concentration, the
images are better fit by a $e^{-r^2}$ (Gaussian) or even $e^{-r^4}$
profile in which the light distribution is both less centrally peaked
and has no extended tail. The isophotes of some ellipticals may be
{\it Boxy-distorted} (\cite{1989A&A...217...35B}) rather than the
elliptical models which have been currently adopted. We will explore
these and alternative models for fitting the continuum of irregular
galaxies in a future paper.

For a point-like stellar image (star or QSO), we need four parameters:
sky background, centroid (x,y), and magnitude. For the extended images
of galaxies, we need at least one extra parameter which measures the
size of the image. Taking into account the image jitter (see
discussion above) and any errors in the PSF, we have found it useful
to adopt a Gaussian profile and to estimate a size parameter even for
the point-like images, to be used as a star-galaxy separation
index. This procedure also takes the stellar image analysis through
the same convolutions as those done for galaxy images, enabling the
likelihood functions to be compared, with some caveats. Errors in the
adopted PSF would appear as an extended residual image following the
model fit. This could make a bright stellar image significantly better
fitted with a model image which includes an extended component. This
is a particularly important issue when attempting to detect underlying
galaxies in QSO images (see \cite{1995ApJ...450..486B}).

\begin{figure}
\plotfiddle{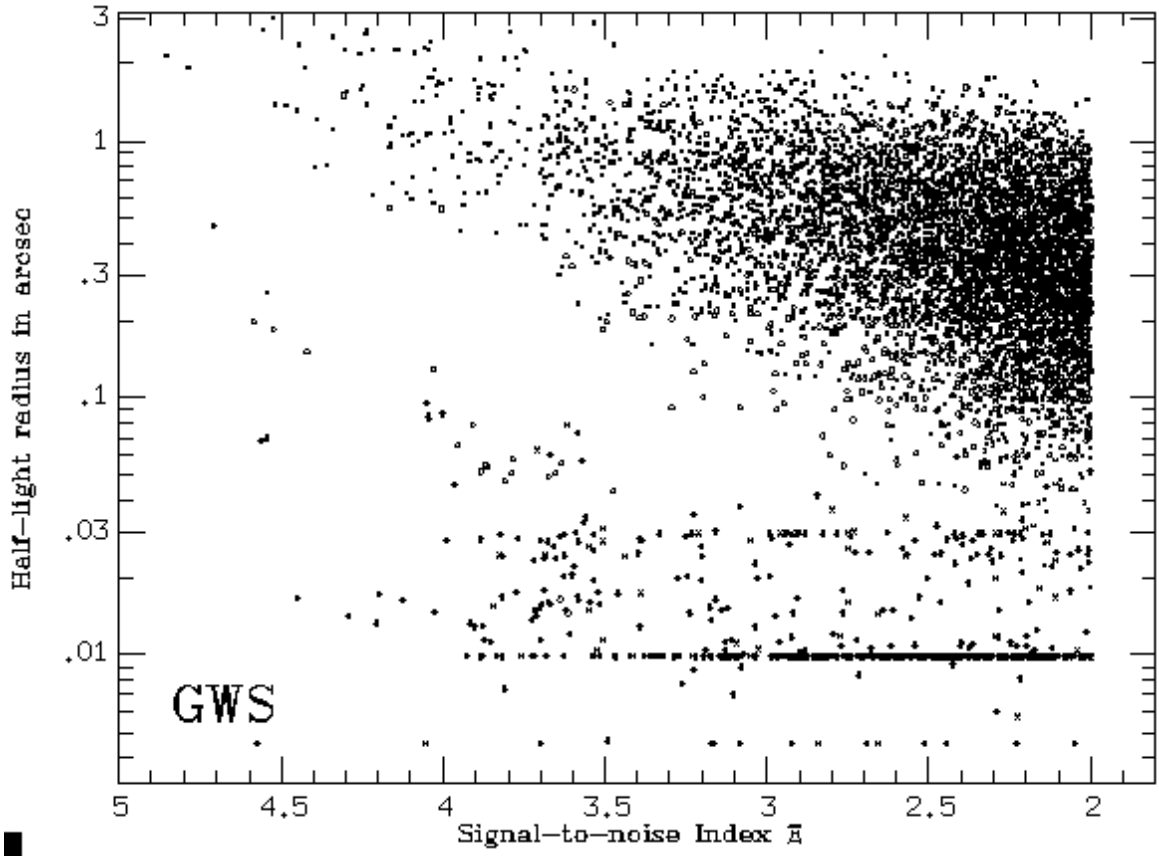}{6.0in}{-90.}{75.}{75.}{-288}{432}
\medskip
\caption{
Plot of half-light radius in seconds of arc as a function $\SNRIL$ for
images in GWS. Symbols as in Fig~22.
\label{fig9}}
\end{figure}

In Figure~\ref{fig9}, we show for the GWS dataset a plot of half-light
radius in seconds of arc as a function $\SNRIL \gtorder 2.0$. For most
stellar images the estimated $\approx 0\farcs02$ or $\HLPX \approx 0.2$ WFPC2
pixel. At brighter magnitudes with $\SNRIL \gtorder 3.5$ we notice some
larger objects (but $\HLPX \ltorder $ 1 pixel) which are very well separated
from the sizes of galaxies. The PSF approximation
adopted in the analysis is insufficient at these bright magnitudes.
They could also be cases of stellar
binaries which are just resolved and which at fainter magnitudes could
contaminate the sample of objects classified as galaxies. 

\section{The parameters}

We describe here the full list of model parameters in the order in
which they are introduced as we increase the number fitted to an
image. These are the intrinsic galaxy model parameters which are
introduced before any PSF convolution or allowance for other
instrumental effects such as (the small amount of) photon scattering
in the CCD before detection.

(1) Sky Background.

The sky background is a very important part of the model estimates. A
bias in the sky estimate could translate to a bias in the estimated
morphology. Unlike for example
\cite{1995ApJ...448..563B,1996ApJ...464...79S} we have therefore
chosen to derive a maximum likelihood estimate for the mean sky
background {\it simultaneous} with the other image parameters. We use
sufficient pixels to ensure that the mean background sky is determined
to an accuracy of 0.5\%. Typical fluctuations of order 1\% are seen in
a single WFC frame. Some of this variation may be caused by the
extragalactic background light (EBL) from faint unresolved
galaxies. Much larger fluctuations are occasionally caused by the
faint halos of nearby images, or by charge transfer problems caused by
bright stars. The estimated sky backgrounds are seen to follow these
variations very well. Sky background is assumed to be flat over the
small region selected for analysis of each object.

In the procedure we have adopted, disk-like or bulge-like model fits
could possibly converge with slightly different sky backgrounds within
the measurement errors. By allowing the sky to vary, we are not
imposing some prior choice of sky background. The error in the sky
background is then properly reflected in the error estimates for the
galaxy parameters and the likelihood ratio used for morphological
classification.
 
(2,3) Centroid

The $(x,y)$ centroid of the model image is in most cases very close to
the centroid of the observed image. The mean errors for $\SNRIL
\approx $ 2.0 and 1.6 are 0.1 and 0.2 pixels respectively. The error
becomes much larger for images fainter than the detection limit
(i.e. $ \SNRIL \ltorder 1.6 $). For the \DPB models we assume the same
centroid for both components. The software does allow an independent
offset for the center of the bulge from that for the disk (parameters
(12,13)), but this has not yet been fully investigated. The extra
degree of freedom resulted in poor convergence in many more galaxies
than in the few which justified it.

(4) Total magnitude.

The adopted magnitude is the analytical total magnitude of the galaxy
model. This estimate has the advantage of not needing an aperture
correction as is required for a fixed aperture or isophotal
magnitude. However, since the magnitude integration is over a smooth
galaxy image, small errors could arise from the fact that the model
may not average properly over bright regions of star formation, for
example. For \DPB models the magnitude is the total for both
components, a quantity better defined than the magnitudes of the
individual components. The magnitudes of the individual disk and bulge
components can be derived using the flux ratio ( $\BTD$ see below).

Note that the total magnitude is integrated theoretically out to
infinity. For disk galaxies practically all (99\%) of the light falls
within 4 half-light radii. However for bulge-like galaxies only 85\%
of the light is within 4 half-light radii, and the model needs to
extend out to 19 half-light radii to contain 99\% of it. The Total
magnitude for an elliptical could therefore be $\approx 15$\% brighter
than when calculated by integration out to a typical isophotal
detection radius, and correspondingly for the $\BTD$.

(5) Half-light radius.

This is the radius within which half the light of the unconvolved
model would be contained if it were radially symmetric (an axis ratio
of unity). For axisymmetric galaxies, this definition is independent
of the observed axis ratio of the galaxy, a parameter which depends on
the intrinsic axis ratio and its inclination to the line-of-sight.

For point-like sources we fit a Gaussian profile with an exponent of
2.0, and the half-light radius is then 0.69 times the scale length.
For disk-like galaxies with a profile exponent of 1.0, it is 1.68
times the exponential scale length. For bulge-like galaxies with a
profile exponent of 0.25, it is the effective radius or 7.67 times the
scale length. For \DPB models it is by definition still the major axis
radius within which half the light of the combined profile is
contained. Like the total magnitude, this is a quantity better defined
than the half-light radii of the individual components.

As a direct consequence of allowing the sky background to be a free
parameter, we need to impose a maximum half-light radius in order to
avoid this parameter from becoming meaninglessly large when a galaxy
with no central concentration is fitted with a disk-like or bulge-like
model. This limit has been set conservatively to equal half the
maximum radius of the region selected for analysis. For $\approx 4$\% of
the galaxy images, the half-light radius converges on this limit, and
those models need to be rejected and flagged for fitting with a less
centrally concentrated model.

From numerical considerations we impose a minimum half-light radius of
a tenth of a pixel on both the major and minor axes of a galaxy. For
\DPB models this minimum is imposed independently for each
component. This assumption does not put any significant constraints on
the axis ratio distribution of galaxies with a half-light radius
larger than one pixel.

The quantity fitted is the logarithm of the half-light radius in
seconds of arc. The half-light radius of the individual disk and bulge
components can be derived using the bulge/(disk+bulge) flux ratio
$\BTD$ and bulge/disk half-light radius ratio (see $\HLF$ below).

(6) Orientation.

The adopted position angle is that of the axis of symmetry of the
galaxy model. Measured in radians in the range $[-\pi/2,+\pi/2]$, this
is set equal to zero when the source is assumed to be azimuthally
symmetric with an axis ratio of unity.

For pre-refurbishment data with a highly asymmetric PSF, the observed
orientation of the image could be significantly different from the
intrinsic orientation of the fitted model. During the minimization
procedure, the angle is measured clockwise from positive Y to positive
X of the relevant CCD. It is then translated into a position angle as
measured clockwise from North towards East using \PAV3 of the HST
attitude (pointing) vectors and the WFPC2 CCD plate-scale distortion
map.

For \DPB models we generally assume that the orientations of the disk
and bulge components are the same. Since the bulge axis ratio is
expected to be close to unity, any difference in orientation could be
expected to be insignificant except in the brightest galaxy
images. The software does allow for a difference in the orientation of
the bulge from that of the disk (parameter(11)), but this too has not
yet been fully investigated.

(7) Axis ratio

This is the ratio of the minor axis half-light radius to that of the
major axis. This parameter has no units and is constrained to be
smaller than unity to ensure proper definition of the major axis. For
\DPB models it is defined independently for each component. If the
axis ratio cannot be shown to be significantly different from unity
then it is held at unity; for the one-component case, the position
angle can then also be dropped as a free parameter. The size of
individual pixels also imposes limits on the ability to usefully
constrain an axis ratio. Note that we adopt a minimum minor axis
half-light radius of 0.1 pixel; i.e. for a Galaxy with a half-light
radius of 0\farcs5 this imposes a lower limit on the axis ratio of
0.02 since WFPC2 has a pixel size of 0\farcs1 . In a few rare cases,
this limit was useful for the prevention of the minimization procedure
from converging on an unrealistically low axis ratio.
This observationally imposed limit could be taken into consideration
in an analysis of the axis ratio distribution, but can practically be
ignored for galaxies with half-light radii larger than 1 pixel.

(8) bulge/(disk+bulge) flux ratio

This is the fractional flux contribution of the bulge-like component
to the (disk+bulge) light ( $\BTD$) in the galaxy image. It has no units
and ranges from zero for pure disk-like galaxies to one for pure
bulge-like galaxies. The ability to estimate this quantity depends on
the integrated signal-to-noise index $(\SNRIL)$ in the image. A larger
$\SNRIL$ is needed to separate out a second component with smaller
fractional contribution to the total light (see Figure~\ref{fig17}). A
second component is only fitted when there is a significant
improvement to the likelihood ratio to compensate for the increased
number of parameters. The definition has used (disk+bulge) rather than
Total to allow for the possible extension of the model parameter set
to a third component such as a central point source (see
\cite{1996ApJ...471L..15S}).

(9) Bulge axis ratio

This is the ratio of the half-light radius of the minor axis to that
of the major axis of the bulge-like component. In \DPB models it is
often a poorly defined quantity when the disk component dominates the
galaxy image, and the ratio is then adopted to be unity. We could not
determine any meaningful relation between the bulge axis ratio and the
disk axis ratio. Such a relation might have been expected if most
disks and bulges have a typical axis ratio and were related by the
common inclination to the observed line of sight. The latter does not
seem to be the case.

(10) The ratio of the half-light radii ($\HLF$) bulge/disk

This is the ratio of the half-light radius of the bulge-like component
to that of the disk-like component. We observed that the logarithm of
this ratio has a weak correlation with the $\BTD$ flux ratio (see
Figure~\ref{fig12}). This correlation has been reported also by
\cite{1985ApJS...59..115K}. For disk-like galaxies this ratio is 
about 0.25 and for bulge-like galaxies the ratio is about 1.6, i.e. on
average, disk dominated galaxies have a disk half-light radius which
is larger than the bulge half-light radius. Such is the case for our
own Galaxy where this ratio is estimated to be about 0.65. However,
there is a factor of 2.5 rms (i.e. one magnitude cosmic scatter) about
the mean relation. It will be interesting to understand this relation
using galaxy structure formation theories like those published by
\cite{1998MNRAS..2XX..XXXM}.

(11) Orientation difference of bulge from disk

 See discussion above on Orientation.

(12,13) Centroid difference of bulge from disk

 See discussion above on centroid.

\section{Optimizing the model fitted}

In brief outline the procedure is as follows:

The initial guess is typically far removed in parameter space from the
final maximum likelihood model fit. At this point it is not useful to
make any judgment about the selection of the model or the parameters
to be fitted. However, testing has shown us that for 70\% of a typical
catalog with $\SNRIL \ltorder 2$, we are never able to fit a significant \DPB
model. These images are analyzed only as stars or pure disks or pure
bulge-like galaxies and the better model is selected. In
Figure~\ref{fig8} we show a histogram of the number of galaxies as a
function of $\SNRIL$. We have highlighted the fraction fitted as \DPB
and the fraction for which we can classify the object as being
significantly disk-like or bulge-like.

We first start with a disk-like model, or if $\SNRIL>2$ we attempt a
10-parameter \DPB model fit. The first fit is a special quick mode of
the minimization routine (modified IMSL 9.2 ZXMIN subroutine that uses
a Quasi-Newton method). This mode of minimization is fairly fast since
it does not attempt to check full convergence. It reaches a point in
the multi-dimensional parameter space which is close enough to the
final answer to investigate the likelihood function and make some
intelligent decisions. These investigations are made after each
minimization, and depend on the number of parameters that were fitted.

The quick mode does not use a higher resolution center (see Appendix).
If a default resolution image had been used for the models, we
investigate whether a high-resolution center will change the
likelihood function. In over 75\% of the tests in a typical catalog
reaching down to the detection limit, the absolute change in the
likelihood function is less than three, which can be considered as
insignificant justification for the introduction of a higher
resolution center. Since we are merging parts of two independently
convolved images, the high-resolution center option is only used when
needed.

If the half-light radius is less than $10^{-2.8+0.5 max(\SNRIL,3.6)}$ arc
seconds, the program branches to test if the object is point-like. As
discussed above, we fit a symmetric 5-parameter Gaussian model to
allow for image jitter and any errors in the PSF. For most images the cut
is at $0\farcs1$ or one WFC pixel with a small increase for the brightest 
images (see fig~9). This test is done for
about 30\% of the objects in the sample, although only about 8\% of
the sample are eventually classified as probable point-like stellar
sources, either stars or quasars. The star-galaxy classification is
based on both the likelihood ratio for the best-fit galaxy model as
well as the evaluated half-light radius for the object, which is
typically 0.2 pixels (equal to the resolution used for the sub-pixel
definition of the PSF).

The next check is to see if a two-component \DPB model, if being
considered, is significantly better than a single-component model with
less parameters. In 60\% of the cases (for $\SNRIL \gtorder 2.0$) the numerical
difference is less than 6 and this is insufficient justification for
the fitting of a \DPB model. If the half-light radius is less than two
pixels we again select a single-component fit. In Figure~\ref{fig8}b
we show the fraction of galaxies as a function of half-light radius
for which we can fit \DPB models and the fraction for which we can
classify the object as being significantly disk-like or
bulge-like. The peak of the distribution for which we can fit \DPB
models is at about 5 pixels, and for obvious reasons we are not able
to do so for galaxies with a half-light radius of less than 2 pixels.
Even if the minimization gave a significant fit for a few of the
latter galaxies, these fits are unlikely to be realistic models of
these extremely under-sampled galaxy images.

For single component galaxies we next check if the axis ratio is
significantly different from unity. If not, then it is set equal to
unity, and a five-parameter symmetric model is fitted to the data. For
all galaxies we fit both a pure disk as well as a pure bulge model,
selecting the better fit model. If the absolute value of the
likelihood ratio is smaller than four, then the classification as disk
or bulge is not significant and these objects are classified as
generic ``galaxy''. If the object had been classified at a longer
wavelength as disk or bulge, then the model output is selected to be
that of the nearest wavelength for which the image was definitively
classified. Otherwise, the model output is based formally on the
likelihood ratio, ignoring the significance of it. For images with a
sub-pixel half-light radius for which the likelihood ratio does not
give a preference between star and galaxy, such objects are classified
merely as ``object''. The star-galaxy separation at sub-pixel
half-light radii needs more detailed investigation, particularly for
the purpose of attempting to isolate an uncontaminated sample of stars
needed for modeling our own Milky-Way Galaxy.

\begin{figure}
\plotfiddle{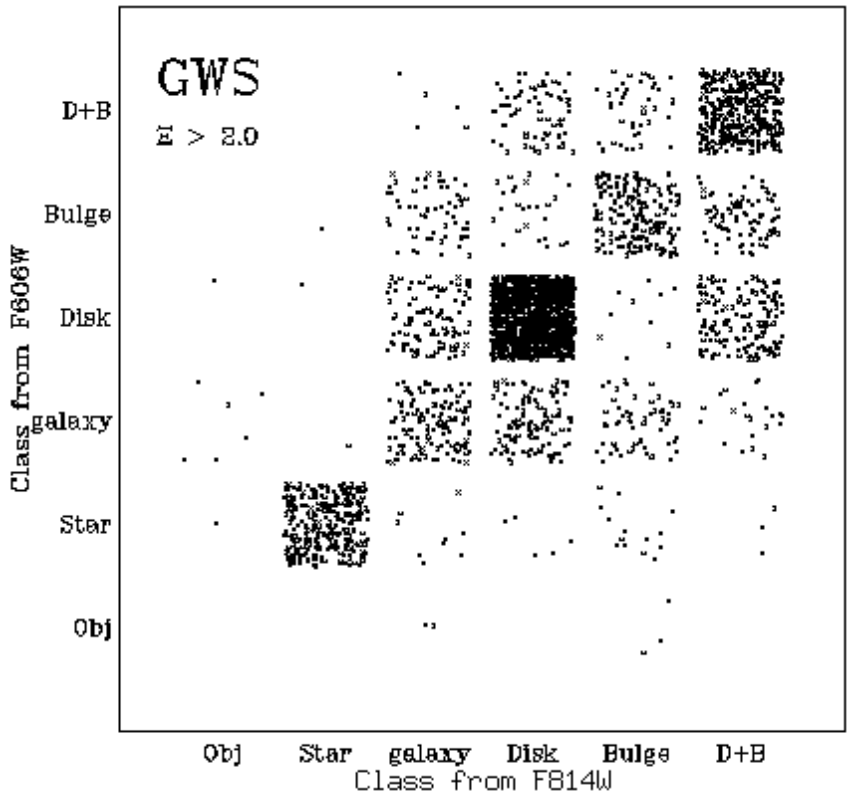}{6.0in}{-90.}{75.}{75.}{-288}{432}
\medskip
\caption{
Comparison of the classification for images in GWS with $\SNRIL > 2.0$
in the two observed WFPC2 filters F606W and F814W.
\label{fig10}}
\end{figure}

The image in each filter is modeled independently since the parameters
in each filter need not be the same. In Figure~\ref{fig10} we compare
the classification of images in GWS with $\SNRIL > 2.0$ in both the
filters F606W and F814W. Most of the objects received the same
classification in the two filters. As expected from Figure~\ref{fig9}
there is very little ambiguity in the star-galaxy classification.

\begin{figure}
\epsscale{.60}
\plotone{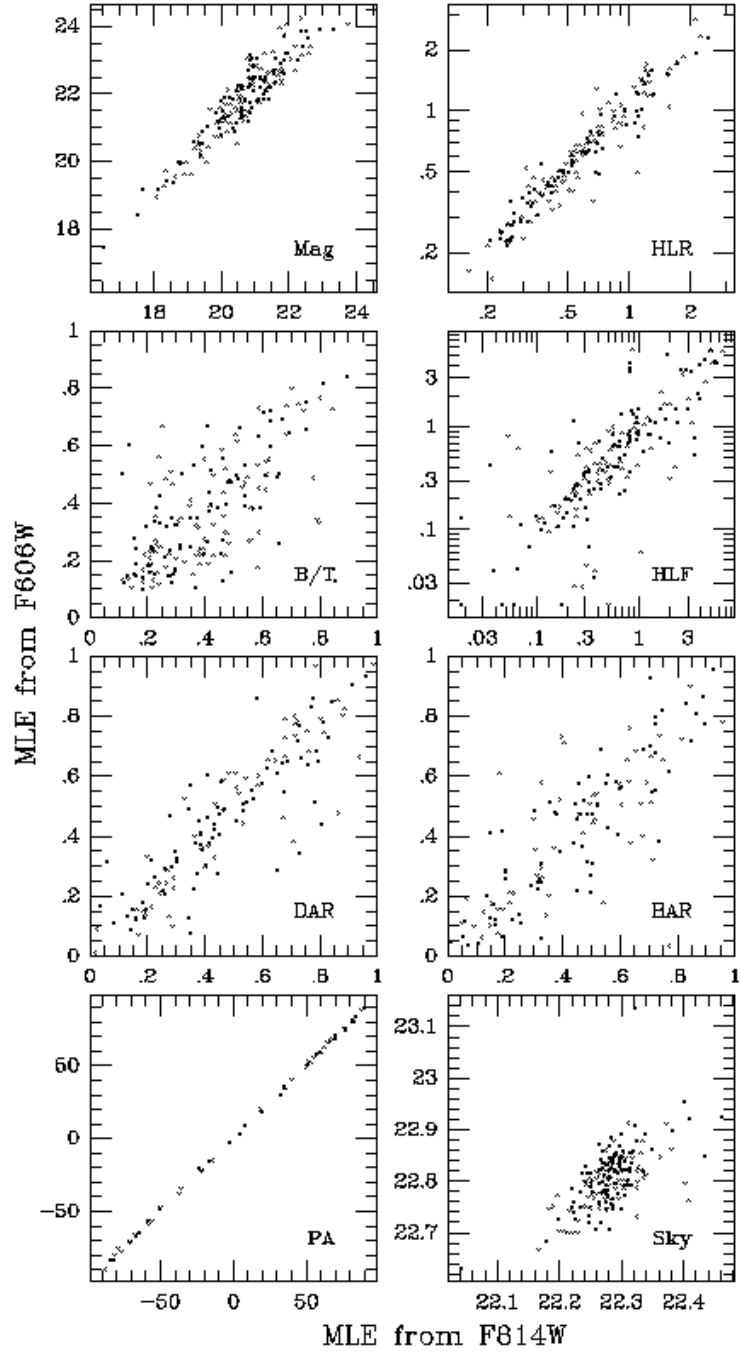}
\medskip
\caption{
Comparison of the parameter estimates for about 150 galaxies in GWS
which got full 10 parameter \DPB MLE fits in both F606W and F814W filters.
\label{fig11}}
\end{figure}

In Figure~\ref{fig11} we compare the parameter estimates for about 150
galaxies in GWS for which there is a full 10 parameter \DPB fit in
both filters and a rms error estimate smaller than 0.5 in $log_e(\HLF )$.
The orientation (PA) is clearly the best defined parameter and this
has proven very useful for studies of weak lensing
(\cite{1996MNRAS.282.1159G}). The deviation for the total magnitude
(Mag) is the color of the galaxy. The half-light radius (HLR) is the
equal in the two filters for most galaxies. The axis ratios [for the
disk components] (DAR) and for the bulge components (BAR) show scatter
mostly from measurement error. The scatter in $\BTD$ flux ratio is
real, and is caused by the different colors of the bulge and disk
components. 

\begin{figure}
\plotfiddle{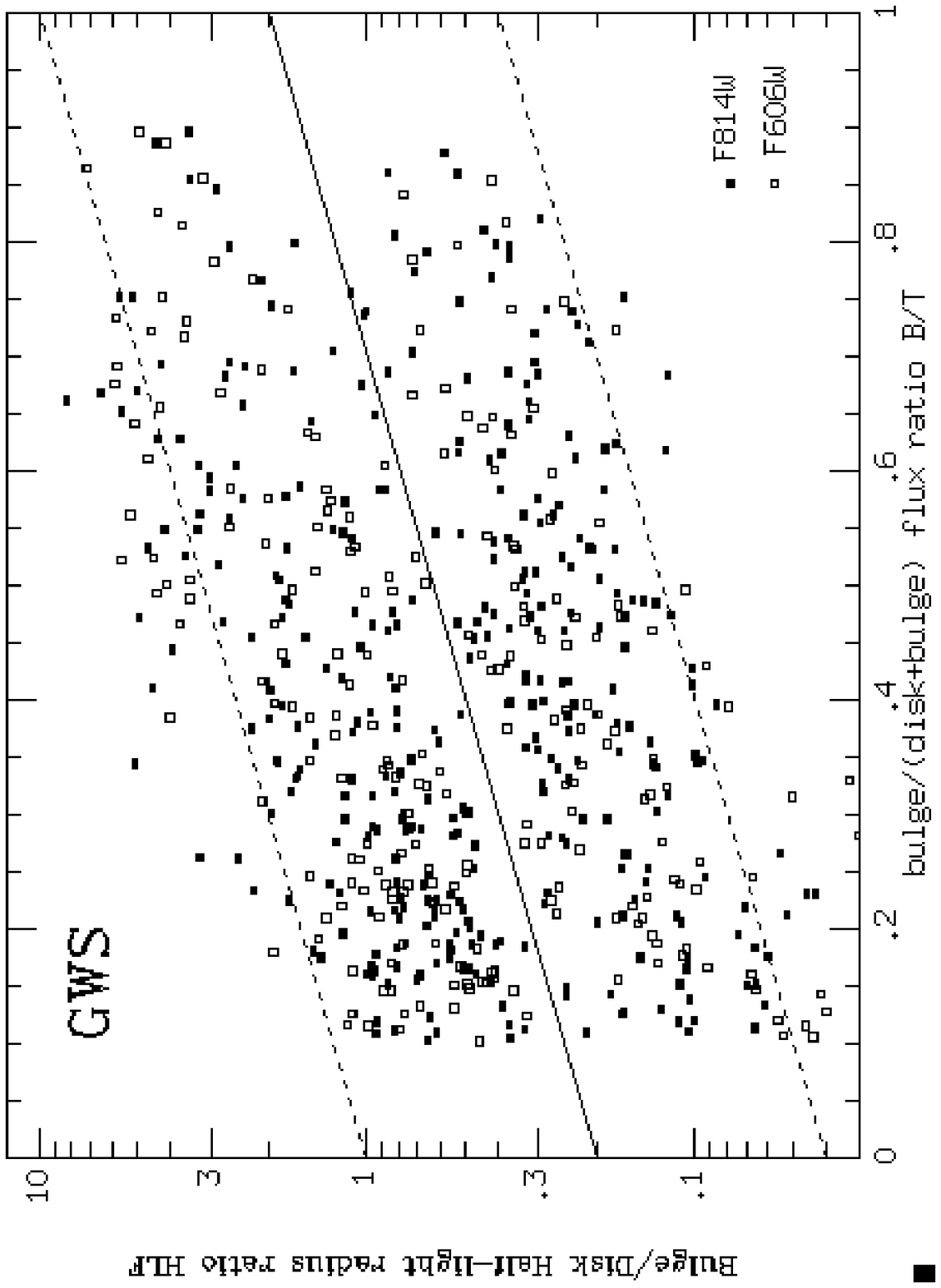}{6.0in}{-90.}{75.}{75.}{-288}{432}
\medskip
\caption{
The observed correlation with 0.4 dex rms cosmic scatter, of the
Bulge/Disk half-light radius ratio ($\HLF$) as a function of $\BTD$
flux ratio for galaxies in GWS. The mean value from relation (solid
line) was adopted if $\HLF$ is unconstrained.
The dashed lines are at $\pm 0.7$ dex.
\label{fig12}}
\end{figure}

For galaxies which demonstrably have two components, i.e. disk and
bulge, the least well defined parameter is $\HLF$ the ratio of the
half-light radii. After a lot of effort, we have optimized an
automated procedure to identify those cases for which a significant
\DPB model can be fitted. We are now able to select and converge (with
over 90\% success) on an unbiased estimate of the ratio of half-light
radii for about half of these cases. The program determines if this
quantity is unconstrained, by searching for a change in the likelihood
as a function of this parameter. If the fainter component contributes
less than 10\% of the light, or if the axis ratio of both components
is unity, then we have generally found this parameter to be poorly
constrained. In Figure~\ref{fig12} we show that the logarithm of this
parameter is a linear function of the $\BTD$ flux ratio with a
correlation coefficient of about 0.5. Bulge dominated galaxies have a
systematically larger Bulge/Disk half-light radius ratio ($\HLF$) than
disk dominated galaxies. However the surface brightness limit for
detection of the fainter component (see Fig.~20) probably contributes
most of the observed correlation. 

The scatter of 0.4 dex rms about the adopted mean relation (solid
line) is equivalent to a cosmic scatter of one magnitude. If in the
preliminary convergence the flux ratio $\BTD < 0.1 $, $\BTD > 0.9 $ or
if the likelihood function was evaluated at extremes $\HLF \pm 0.7$ dex
showed that the ratio of the half-light radius ratio was not
constrained by the data, it is held fixed at the nominal value derived
from the empirical relationship

   $$ \log_{10} ( \HLF ) = -0.7 + \BTD $$

Such relationships, although needed to facilitate convergence of the
model fits at fainter magnitudes, are at best a rough approximation.
However, when a parameter is unconstrained and the errors become
comparable to the expected range of parameter space, this assumption
does not significantly change the estimates of better defined
parameters. The justification for the application of such a
relationship is that it helps the routine to converge on a better
defined minimum.

The program may also choose to fix the bulge axis ratio, or less
frequently, the disk axis ratio at unity, if either of them is
determined to be not different statistically from unity.

\section{Estimated errors of parameters}

The covariance matrix is the inverse of the Hessian i.e. the
second-order derivatives evaluated at the peak of the likelihood
function. When it is normalized to have unit diagonal elements, the
cross-terms then give the correlation coefficients between the
estimated model parameters. If the cross-correlation terms are not
large, we can expect to derive reliable error estimates for the
parameters from the diagonal elements. The parameters were selected to
try to minimize the covariance between the fitted parameters described
above.

In MLE theory, if the image being modeled is the same as the simple
model assumed, then the parameter estimates and associated errors will
be unbiased. However, real galaxy images which are well resolved are
more complex than the simple axisymmetric image models that are
assumed for MLE. The effects of spiral arms and bars on the parameter
estimates are complicated and difficult to quantify using simulations.
In general, we can expect that, given a sufficiently large sample, the
cosmic dispersion caused by image peculiarities will be averaged out.

\begin{figure}
\epsscale{.75}
\plotone{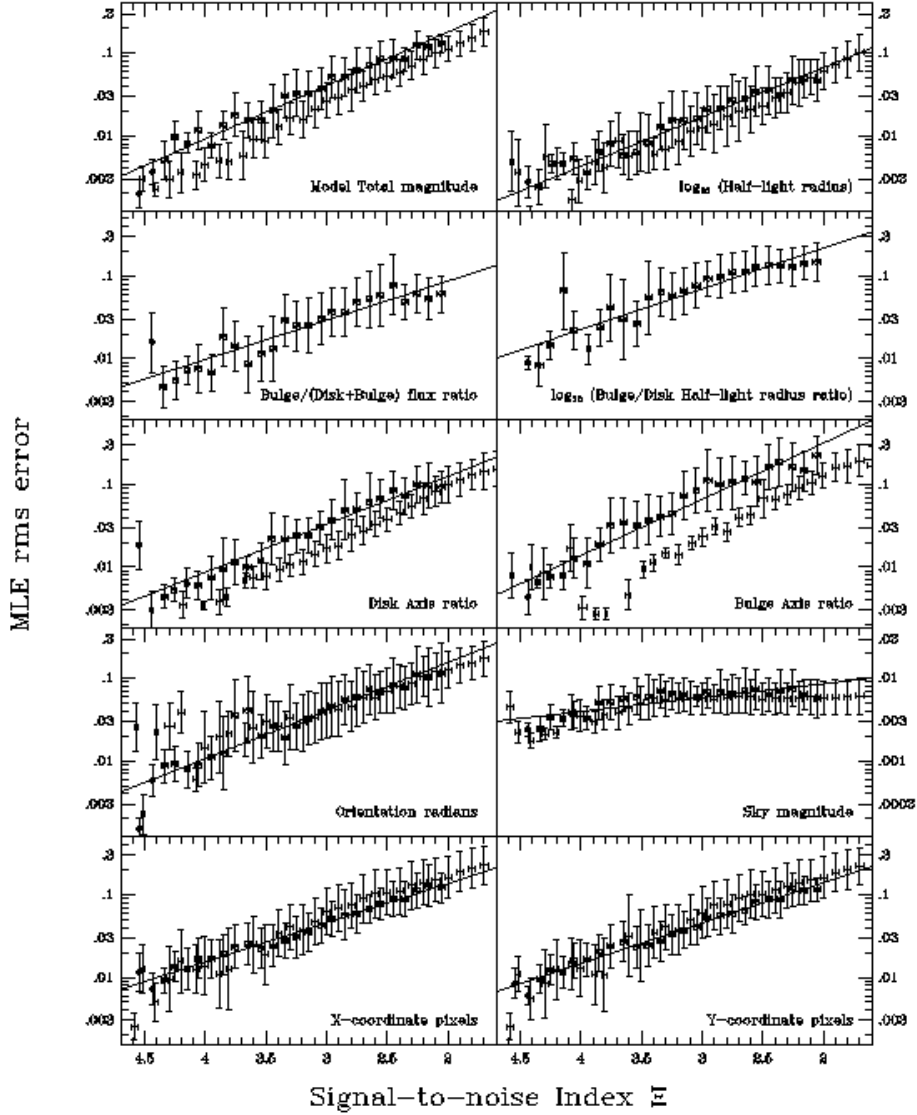}
\medskip
\caption{
The running mean of estimated rms errors for parameters as
a function of $\SNRIL$ for single component and \DPB MLE model fits
for galaxies in GWS. See text for details.
\label{fig13}}
\end{figure}

In Figure~\ref{fig13} we illustrate a running mean of rms errors for
parameters as a function of $\SNRIL$. To first order, the logarithm of
the rms error appears to increase linearly with $\SNRIL$. The errors
for single component and two component \DPB fits are illustrated
independently: in general, the latter errors are larger. There are a
few points to notice. Firstly the sky error, of order 0.005 magnitude,
is practically independent of $\SNRIL$ and is defined by our choice of
the number of sky pixels to include in the MLE. The orientation and
centroid position, which were held the same for both components, show
no significant increase in error than a single component fit at the
same $\SNRIL$. The errors in the bulge axis ratio are much larger for
the two component fits. Since the rms of a random distribution between
0.13 and 1.00 is 0.25, rms errors larger than $\approx 0.1$ convey little
useful information about the axis ratio. This occurs at a $\SNRIL$ of
1.93 and 2.12 for single component disk-like and bulge-like galaxies
and $\SNRIL\approx 2.25$ and 2.72 for the disk and bulge components in \DPB
model fits. The $\BTD$ errors do not become larger than 0.1 since a
two-component model would not be significant if they did. The error in
half-light radius is given in $log_{10}$ units. The error is 0.1 dex
or 26\% at $\SNRIL$ values of 2.15 and 2.37 for single component and
two-component models respectively. The $\HLF$ ratio, given in
$log_{10}$ units, is clearly the worst constrained parameter,
requiring $\SNRIL \gtorder 3.37$ for the expected error to be less than 0.1
dex or 26\% rms.

The HDF superstack consisted of eleven individual HDF field pointings,
and we can therefore use these to test the MLE method. We compare the
MLE results for the HDF super-stack with the MLE results of the
independent fits to the images of the same galaxies in each of the 11
sub-stacks. We limit the comparison to those galaxy images where the
output from the sub-stacks resulted in the same morphology
classification as that from the super-stack in the same filter. In the
fits to all of the sub-stacks, we used the same object definition mask
(see appendix) as that derived from object detection in the
super-stack, together with the appropriate shifts.

\begin{figure}
\epsscale{.60}
\plotone{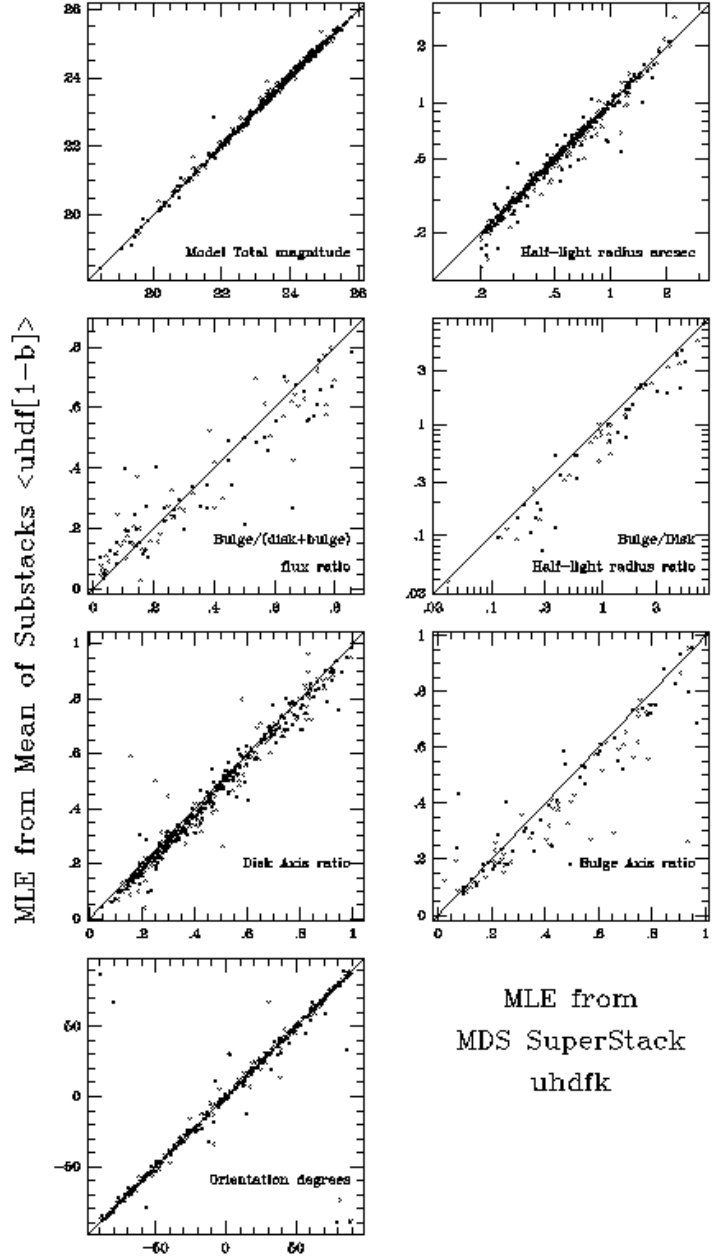}
\medskip
\caption{
Comparison of parameter derived from our HDF super-stack (X-axis) with
the weighted mean of the individual sub-stacks (Y-axis). See text for
discussion of small systematic biases observed.
\label{fig14}}
\end{figure}

In Figure~\ref{fig14} we compare the MLE parameters derived from the
super-stack with the weighted means from the individual sub-stacks. We
notice a small systematic bias: the axis ratios in the super-stack are
slightly rounder and the half-light radii slightly larger, with a
slightly larger $\HLF$ ratio. Our adopted approach to stack after
shifting by closest integer number of pixels modifies the appearance
of the peaked bulges. It will be instructive to see if the process of
``drizzling'' (\cite{1997STScI.CT3..518F}) and stacking with sub-pixel
shifts helps to remove this effect completely. The errors in flux
values of pixels in drizzled images are not independent and to use our
MLE approach, the covariance error matrix for each pixel needs to be
included in the evaluation of the likelihood function. Software to do
this has yet to be developed. For the brighter galaxy images in the
HDF it is probably better to use a weighted mean estimate of the
galaxy parameters from the individual HDF sub-stacks rather than the
MLE values derived for the super-stack.
Although the image bias in our HDF super-stack is disappointing, it
does show the power of MLE estimates to be sensitive to the true
nature of the images analyzed. Of course, all these problems can be
avoided by not stacking the images at all and, instead, by summing the
likelihood over the individual images (\cite{1994STScI.RT2..333R}).
This latter approach, however, is computationally impractical as yet.

Figure~\ref{fig13} allowed us to easily estimate an expected error for
a given $\SNRIL$. If the error estimate from inverting the Hessian is
significantly smaller then it is unlikely to be real. This could
happen for many reasons. There could be a sufficient covariance
between parameters to make the diagonal only a small part of the
error. The non-axisymmetric features of the galaxy image could have
made a sharper dip in the likelihood function. The expected error
could in fact be built into the evaluation of the Hessian at the peak
of the likelihood function in order to pass over any sharp dips in the
function. However, these relationships had not been derived at the
time of the 1996-98 MDS pipeline processing. We find that a reasonable
compromise for the current (October 1998) version of the database is
to adopt a nominal expected error of half a magnitude brighter object
if that is larger than the MLE error estimate from the Hessian. We
find this is appropriate for all parameters except the orientation
parameter, for which the original error estimates appear to be
good. The orientation is not correlated with any of the other image
parameters.

\begin{figure}
\epsscale{.75}
\plotone{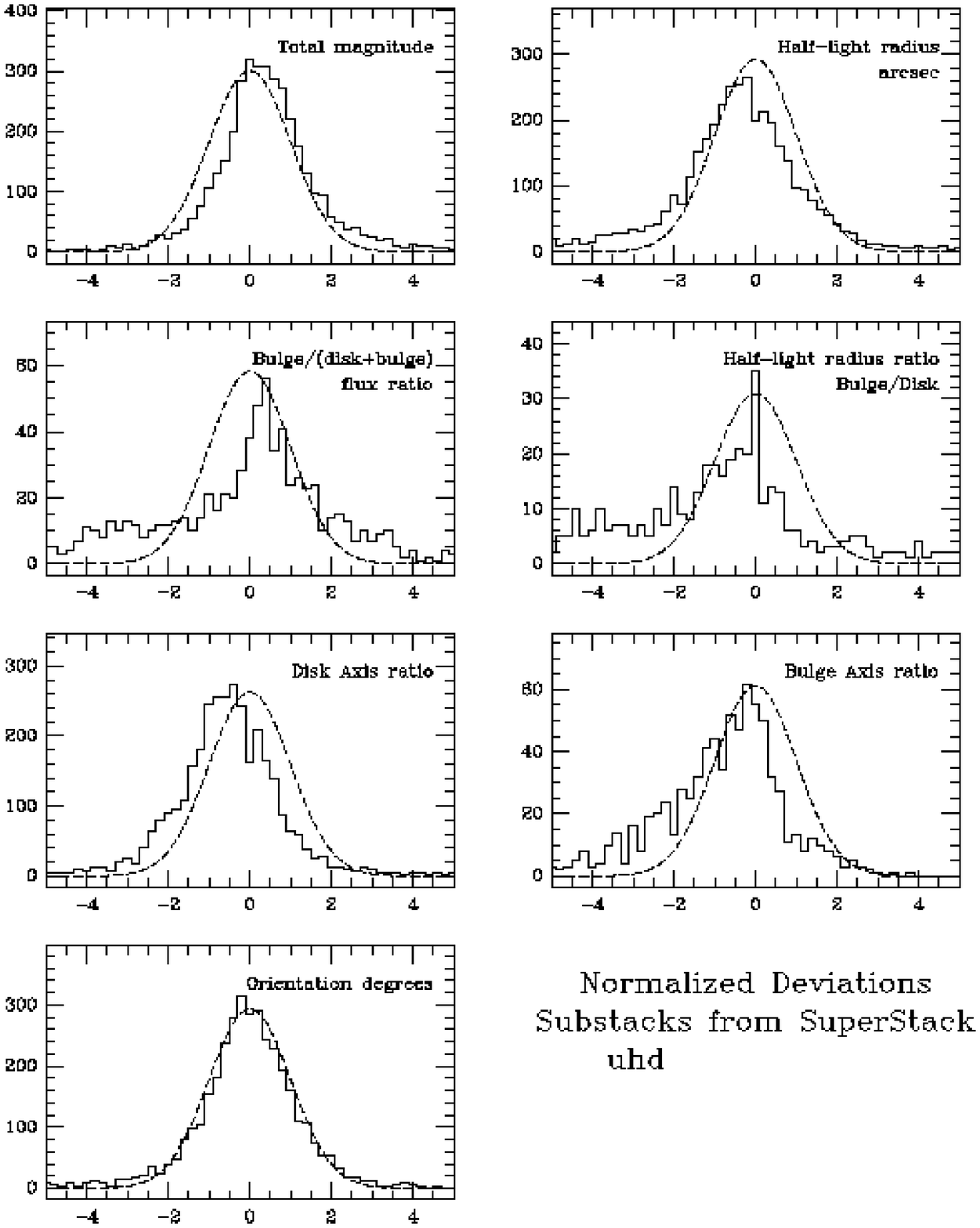}
\medskip
\caption{
A histogram of the normalized deviations of the parameter
estimates derived from the individual HDF sub-stacks from the value derived
from the super-stack is compared with the expected standard normal 
distribution. See text for details.
\label{fig15}}
\end{figure}

In Figure~\ref{fig15} we show a histogram of the resulting normalized
deviations of the parameter estimates evaluated in the individual
sub-stacks from the value derived from the super-stack, and we compare
the results with the expected standard normal distribution. The small
bias caused by stacking the parameter estimates discussed above is
clearly emphasized. We see a significant tail larger than normal for
the $\BTD$ flux ratio and for the $\HLF$ ratio because of the residual
covariance in these parameters. The overall accuracy of the MLE
parameter error estimates seems reasonable if we recognize that the
simple galaxy model fitted does not include the structural detail seen
in the real galaxy images at brighter magnitudes.

\section{Selection effects effects due to Ockham's razor}

Since the adopted procedure fits the minimum number of parameters
which are required to get a best MLE fit which is statistically
significant, the parameter estimates reflect that decision.

\begin{figure}
\epsscale{.75}
\plotone{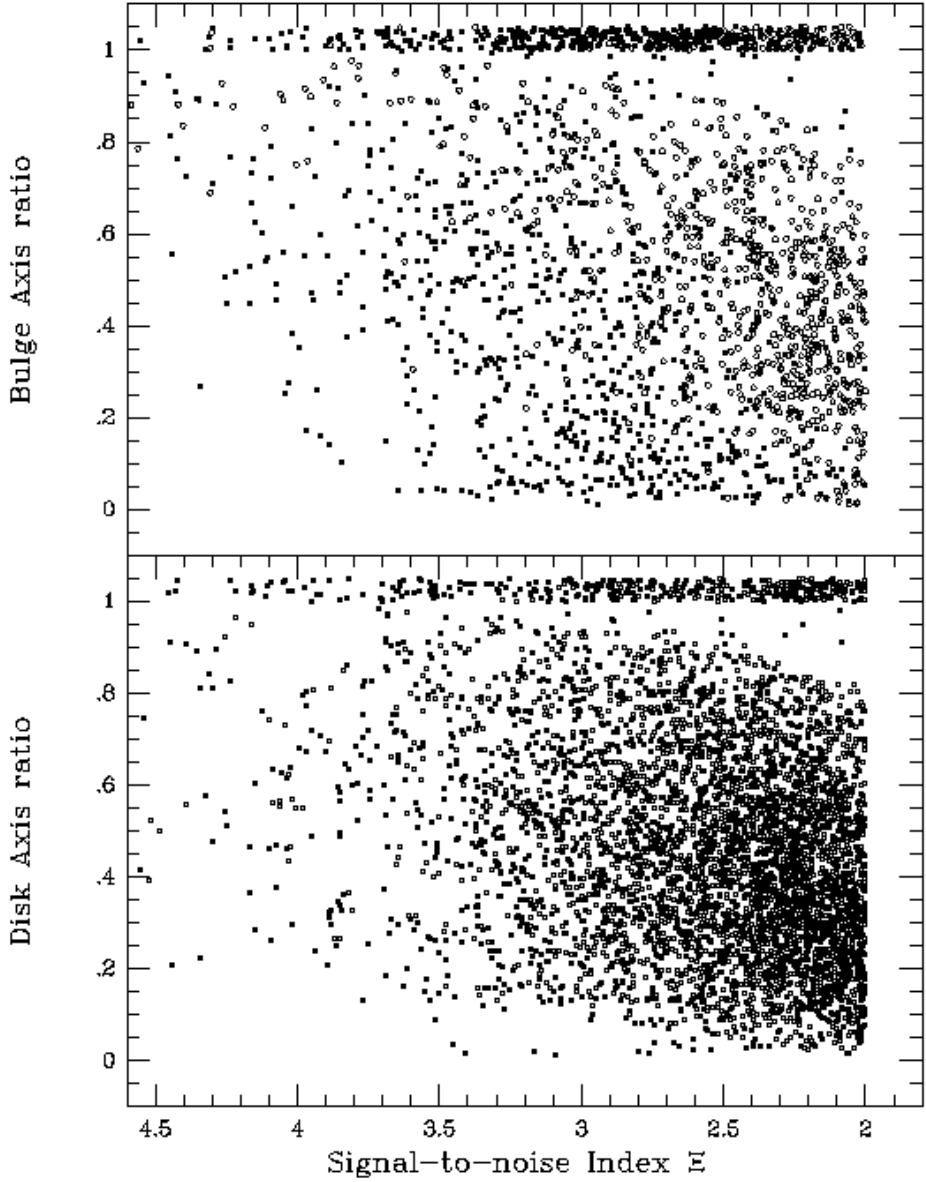}
\medskip
\caption{
The Distributions of estimated disk and bulge axis ratios as
a function of $\SNRIL$ for galaxies in GWS. For illustration the face-on (axis
ratio = unity) case has been distributed randomly in the finite range
[1.00,1.05] outside the fitted range [0.01,1.00].
Symbols as in Fig~22. See the text for details.
\label{fig16}}
\end{figure}

In Figure~\ref{fig16} we show the distributions of disk and bulge axis
ratios as a function of $\SNRIL$. For illustration, the face-on case
(axis ratio = unity) has been distributed randomly in the finite range
[1.00,1.05] outside the fitted range [0.01,1.00]. The disk axis ratio
appears to be randomly distributed within the range [0.10,1.00] for
$\SNRIL$ brighter than $\approx 3.0$. As images get fainter the axis
ratios close to unity are found to be insignificantly different from
unity. For example at $\SNRIL \approx 2$, axis ratios in the range [0.8,1.0]
get set equal to unity, thus removing two parameters from the MLE
fit. The same increase in errors produces a scattering of the observed
axis ratios below 0.10. The bulge axis ratios show a similar
distribution except that they are expectedly larger than the disk axis
ratios. We also notice a number of small bulge axis ratios which are
spurious and caused by barred galaxies which have not been properly
included in the current MLE models.

\begin{figure}
\plotfiddle{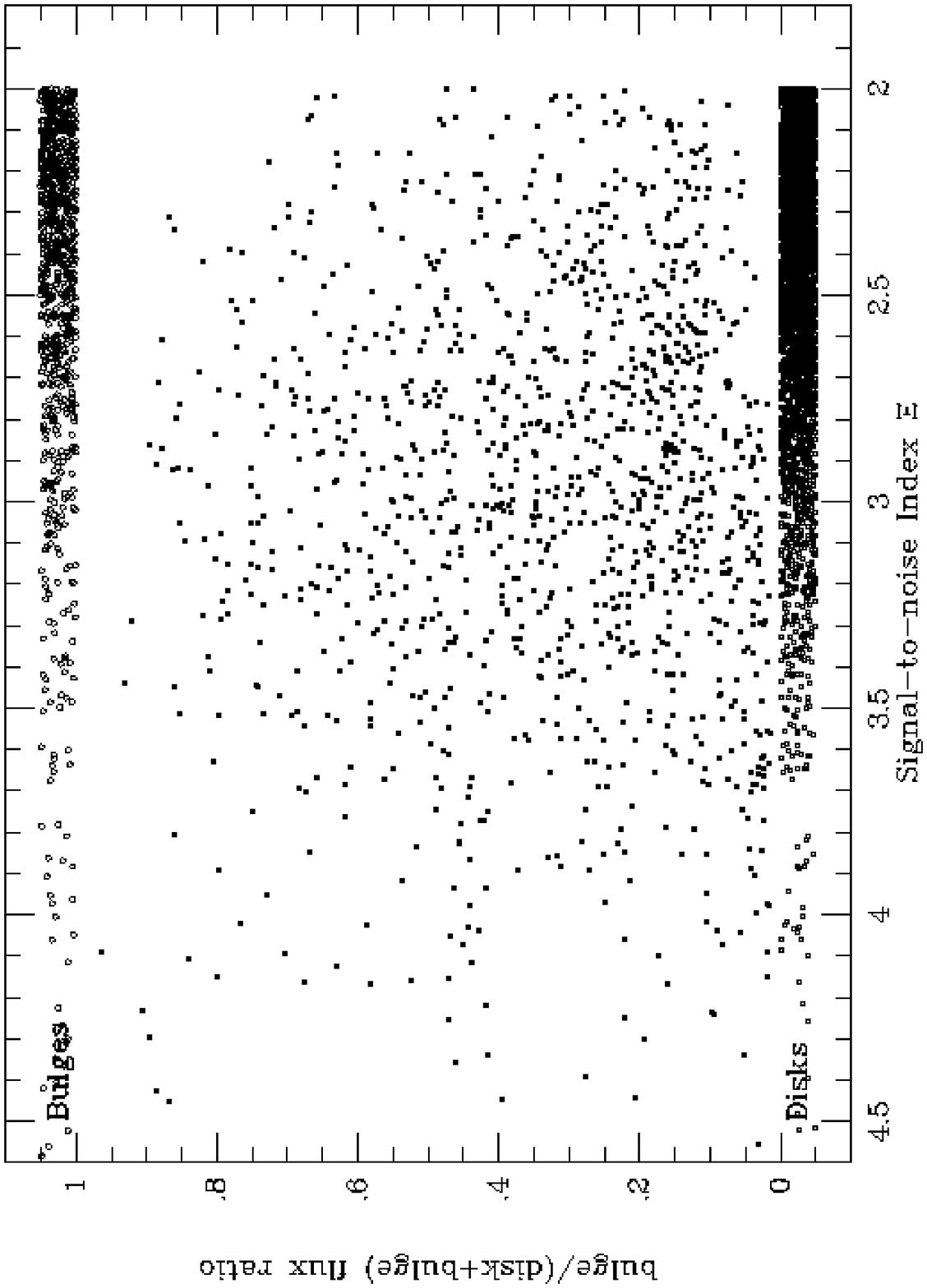}{6.0in}{-90.}{75.}{75.}{-288}{432}
\medskip
\caption{
The Distributions of $\BTD$ flux ratio as a function of $\SNRIL$ for
galaxies in GWS. For illustration, single component fits as pure disks
$\BTD=0$ and pure bulges $\BTD=1$ have respectively been distributed
randomly in the finite ranges [-0.05,0.00] and [1.00,1.05] outside the
fitted range [0.00,1.00]. See text for details.
\label{fig17}}
\end{figure}

In Figure~\ref{fig17} we show the distributions of $\BTD$ flux ratio
as a function of $\SNRIL$. For illustration, single component fits as
pure disks $\BTD=0$ and pure bulges $\BTD=1$ have respectively been
distributed randomly in the finite ranges [-0.05,0.00] and [1.00,1.05]
outside the fitted range [0.00,1.00]. The $\BTD$ flux ratio is
distributed within the fitted range [0.00,1.00] for $\SNRIL$ brighter
than about 3.0, with the understood excess of disk like galaxies. As
images get fainter, ratios close to zero and unity are not observed
since these galaxies do not show a significant second component. The
disk component in ellipticals is `lost' before small bulges are lost
in disk-like galaxies. For example at $\SNRIL\approx 2$, the observed $\BTD$
flux ratios are in the approximate range [0.1,0.6]

\begin{figure}
\plotfiddle{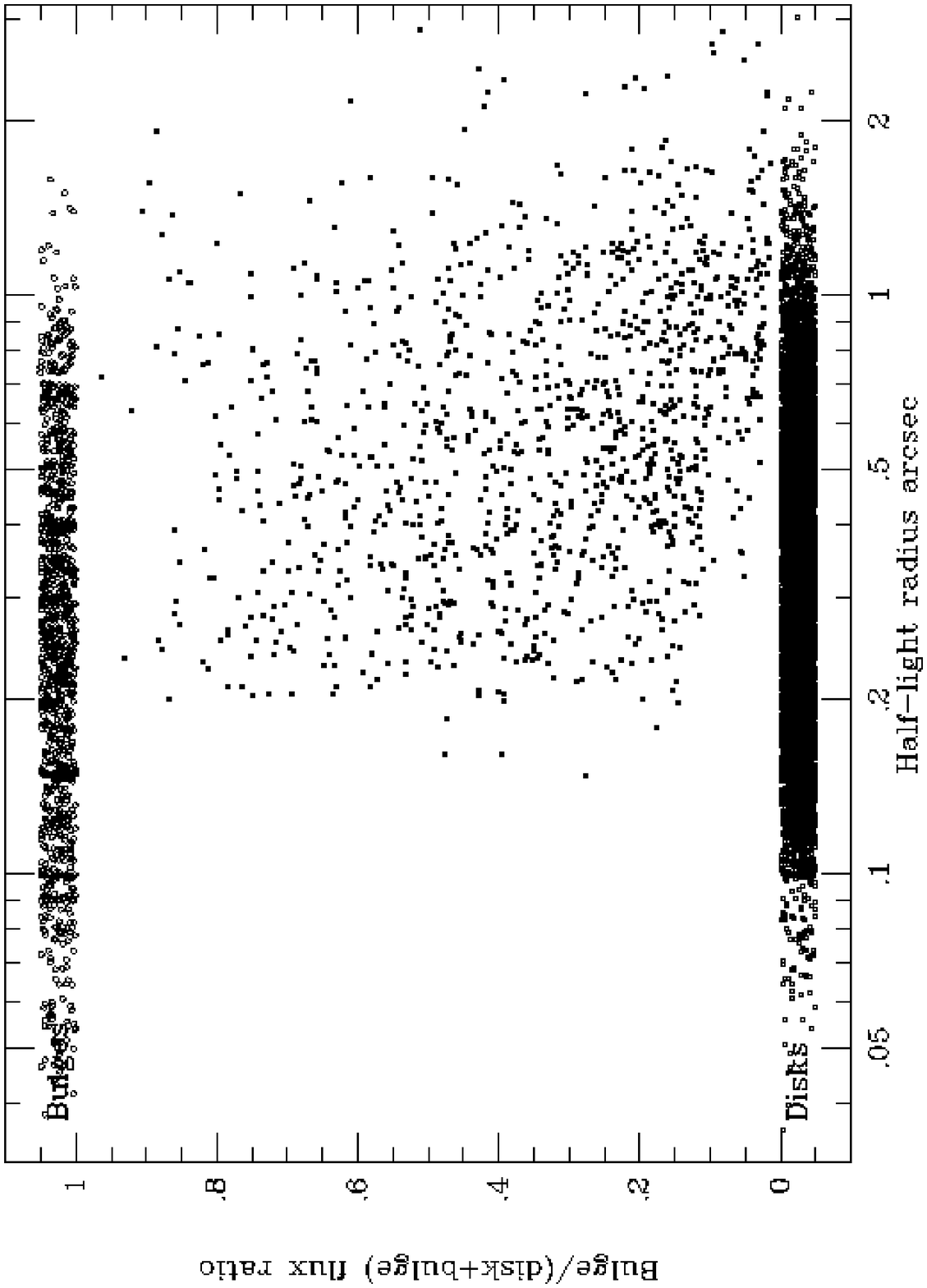}{6.0in}{-90.}{75.}{75.}{-288}{432}
\medskip
\caption{
Distributions of $\BTD$ flux ratio
as a function of half-light radius for galaxies in GWS. The distribution for
single component fits are as in fig~17. See text for details.
\label{fig18}}
\end{figure}

In Figure~\ref{fig18} we show the distributions of $\BTD$ flux ratio
as a function of half-light radius. The distribution for single
component fits are as those in Figure~\ref{fig17}. As galaxy images
get smaller, the $\BTD$ flux ratios close to zero and unity are not
observed since for these galaxies a significant second component
cannot be resolved. Not unexpectedly, small bulge components in
spirals can be inferred to have been lost from the MLE models of
galaxies with half-light radii of a few pixels.

\begin{figure}
\plotfiddle{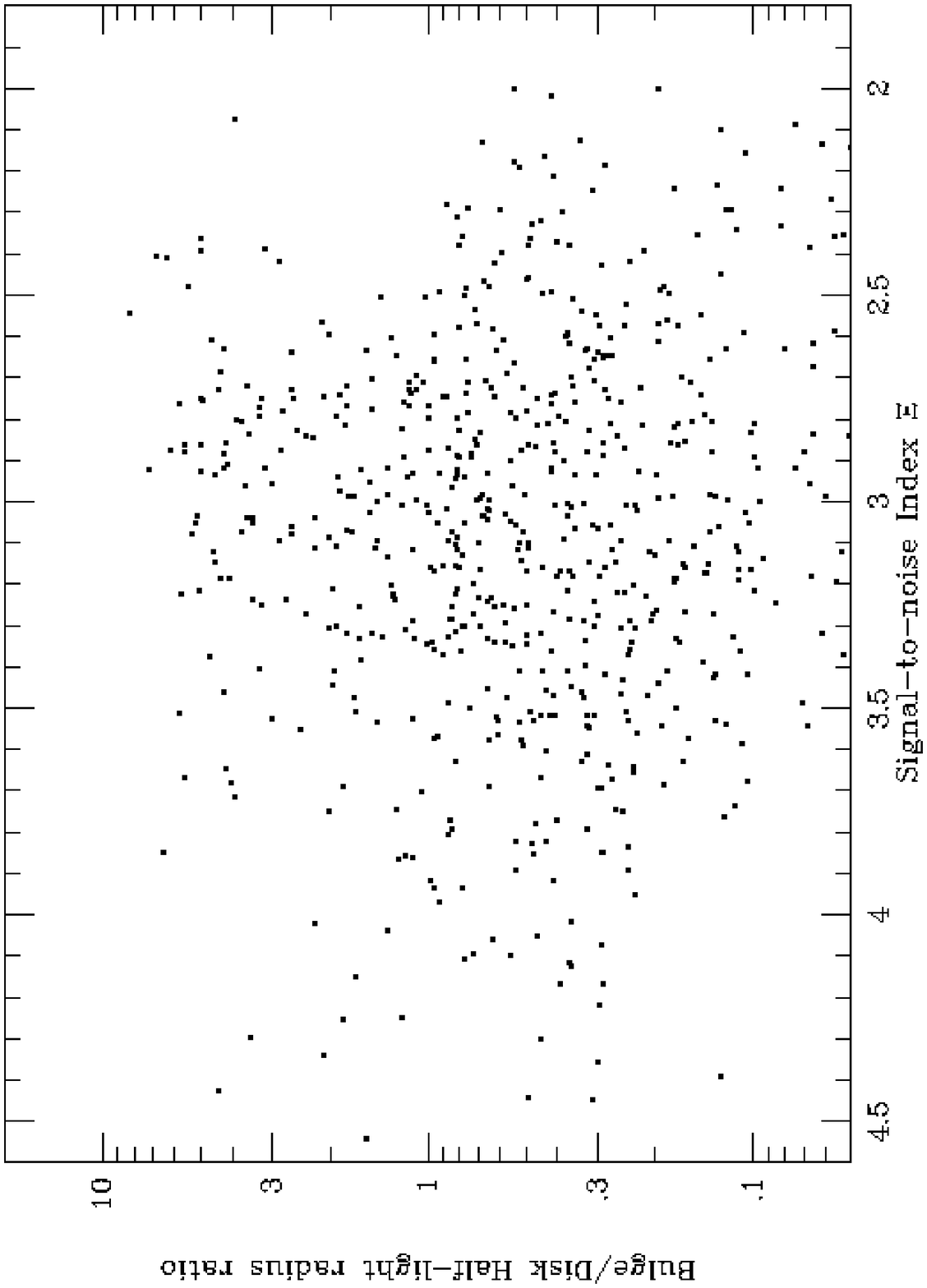}{6.0in}{-90.}{75.}{75.}{-288}{432}
\medskip
\caption{
Distributions of Bulge/Disk half-light radius ratio ($\HLF$) as a function of
$\SNRIL$ for galaxies in GWS. See text for details.
\label{fig19}}
\end{figure}

In Figure~\ref{fig19} we show the distributions of the $\HLF$ ratios
as a function of $\SNRIL$. For $\SNRIL$ brighter than about 3.0 the
ratio is seen to be distributed over a wide range. As images get
fainter than those corresponding to $\SNRIL\approx 2.4$, the MLE routine does
not estimate ratios larger than unity. This is because, as seen in
Figure~\ref{fig17}, the MLE program does not resolve disk-like
components in faint galaxy images dominated by bulges.

We now look at the mean surface brightness within the central
half-light radius ellipse. This has a constant magnitude offset from
the central surface brightness of $-1.12463$ mag for disks and $-6.18126$
mag for bulges. The central surface brightness is a commonly quoted
quantity, independent of axis ratio and inclination for our simple
galaxy models. The advantage of discussing mean surface brightness
here is that we find that the limiting mean surface brightness for
morphological classification is a quantity which is about the same for
disk-like and bulge-like components of the galaxy.

\begin{figure}
\plotfiddle{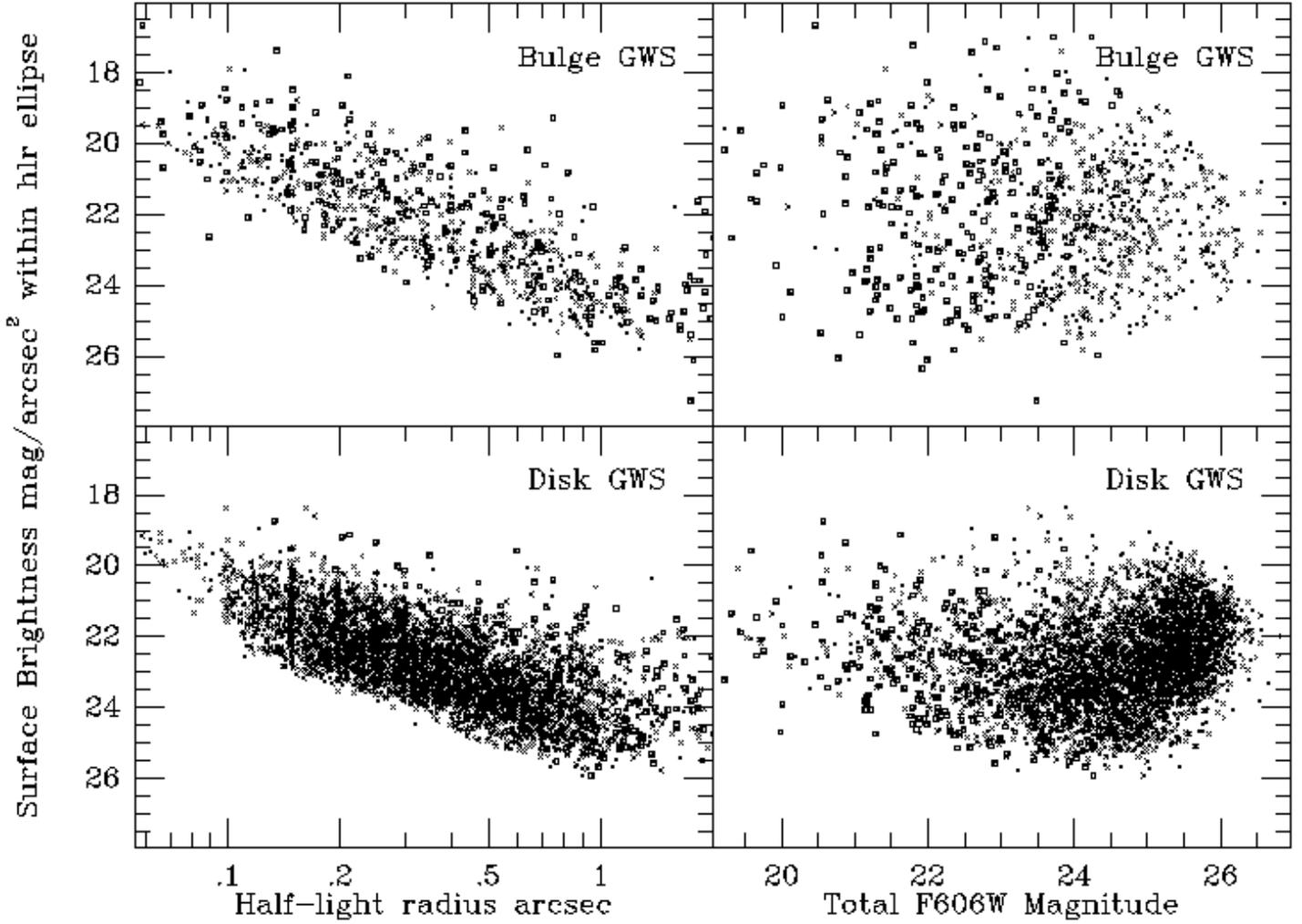}{6.0in}{-90.}{75.}{75.}{-288}{432}
\medskip
\caption{
Mean surface brightness within the half-light radius ellipse as a
function of major-axis half-light radius (left) and total magnitude
(right) plotted independently for bulge (top) and disk (bottom) components.
Cross symbol used for galaxies fitted with single component Model.
\label{fig20}}
\end{figure}

On the left-side of Figure~\ref{fig20} we show the mean surface
brightness within the half-light radius ellipse as a function of the
estimated major-axis half-light radius. In the case of \DPB models,
each of the components are considered separately. There is clearly a
limiting magnitude for morphological classification which appears to
be the same in each case, i.e. independent of whether it was a
component of a \DPB galaxy model, or a single component. We illustrate
this for the GWS 4-stack images of 2800 seconds in F606W. A very
similar graph is seen for F814W. There appears to be a slight
numerical bias for MLE to converge on integer or half-integer
half-light radii at the smaller values. This bias is presumably caused
by our attempt to merge in a high-resolution center (see appendix).

On the right-side of Figure~\ref{fig20} we show the same surface
brightness estimates within the half-light radius ellipse as a
function of the total magnitude of the galaxy. Within the half-light
radius ellipse of a galaxy or component, morphological classification
can be done to a limit in surface brightness which is independent of
the total magnitude of the galaxy up to certain magnitude
limits. These two magnitude limits will be very useful as simple
selection criteria in future models used to interpret the observed
$\BTD$ distribution of galaxies and surface brightness dimming
for cosmology.

\section{Results}

Preliminary versions of the MDS catalog have been the source of
many scientific investigations: see, for example the papers on the 
size - redshift relation (\cite{1994ApJ...434L..55M});
angular size evolution (\cite{1995ApJ...441..494I,1996MNRAS.282.1247R,1997MNRAS.288..200R,1998MNRAS.293..157R})
axis ratio distribution (\cite{1995ApJ...445L..15I});
weak gravitational lensing (\cite{1996MNRAS.282.1159G});
luminosity functions of elliptical galaxies (\cite{1996ApJ...461L..79I});
morphological classification (\cite{1996MNRAS.281..153O,1997ApJ...476..510N,1997ApJS..111..357N,1999ApJ...510...82I});
galaxy interactions and mergers (\cite{1997ApJ...480...59N});
compact nuclei (\cite{1996ApJ...471L..15S})
the HST MDS cluster sample (\cite{1998AJ....116.2644O})
and a study of high-redshift clusters (\cite{1998AJ....116..584L}).

The catalog used in these analyses was mostly based on the star, disk or
bulge model that best fit each object. Most of the previous analyses
can be repeated on the new catalog and refined using the \DPB models
for the brighter sample. We do not, however, expect any significant
changes to these previously reported results.

It is especially interesting to look at results on the two observables
which have not been previously measured for large numbers of galaxies,
especially in the magnitude range observed here, viz. the $\BTD$ flux
ratio and the Bulge/Disk half-light radius ratio ($\HLF$). In fact we
need to apply the same procedure to a large sample of bright nearby
galaxies like those from the SLOAN digital sky survey
(\cite{1998AJ....116.3040G}) in order to establish the behavior of
these parameters on galaxies in the local universe.

\section{Surface Brightness}

\begin{figure}
\plotfiddle{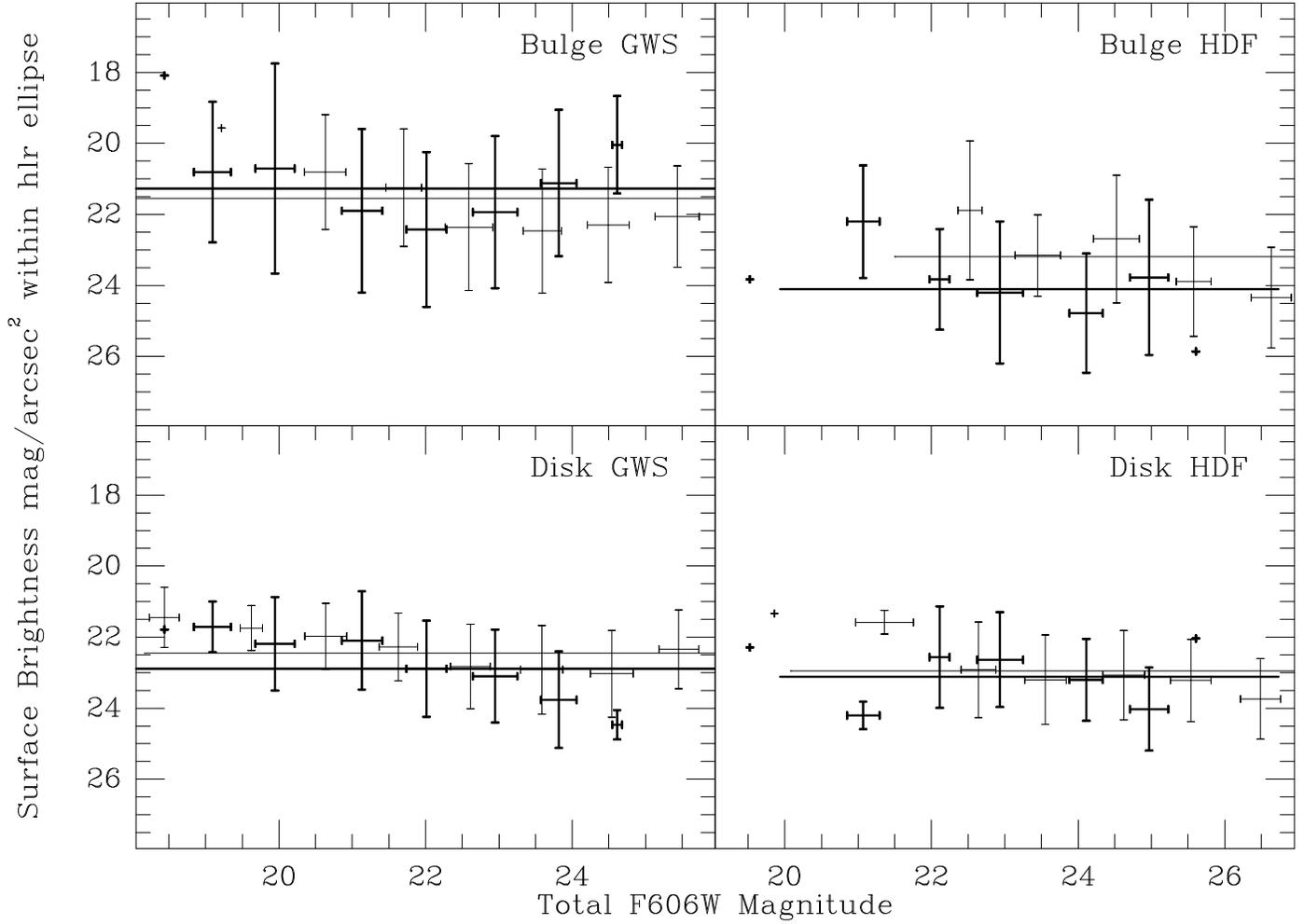}{6.0in}{-90.}{75.}{75.}{-288}{432}
\medskip
\caption{
The running mean of surface brightness as a function of total
magnitude for galaxies in GWS (left) compared with HDF (right).
Components from \DPB models illustrated with thicker line follow same
sequence as galaxies fitted with single component models.
See text for discussion.
\label{fig21}}
\end{figure}

We have made plots similar to Figure~\ref{fig20} for much deeper
observations such as the Hubble Deep Field (HDF). In
Figure~\ref{fig21} we show a running mean of surface brightness as a
function of total magnitude for the GWS galaxies on the left-side and
compare it with those estimated for the HDF on the right-side. This
graph illustrates Freeman's result for disk galaxies
(\cite{1970ApJ...160..811F}); the mean is the same, indicating that
the observed distribution of surface brightness is intrinsic to the
galaxies, with a cosmic dispersion of only about 1-magnitude. The
expected trend of surface brightness dimming as the mean redshift
increases for galaxies with fainter total magnitude is also seen.
Correcting by $-1.12$ mag., we
estimate that the mean central surface brightness of disk galaxies is
20.6, 21.4 mag in F814W, F606W for the GWS and 21.0, 21.8, 22.4 mag in
F814W, F606W, F450W for the HDF. It is interesting that the mean
surface brightness is the same for galaxies fitted as pure disk, as it
is for the disk component of \DPB galaxies. For bulges the scatter
appears to be very much larger and our observations in GWS do not
reach the limiting mean surface brightness bulges. Consequently, the
mean for bulges in the HDF is about 2-magnitudes fainter than for
bulges in the GWS fields. 

\section{Galaxy Color}

\begin{figure}
\epsscale{.75}
\plotone{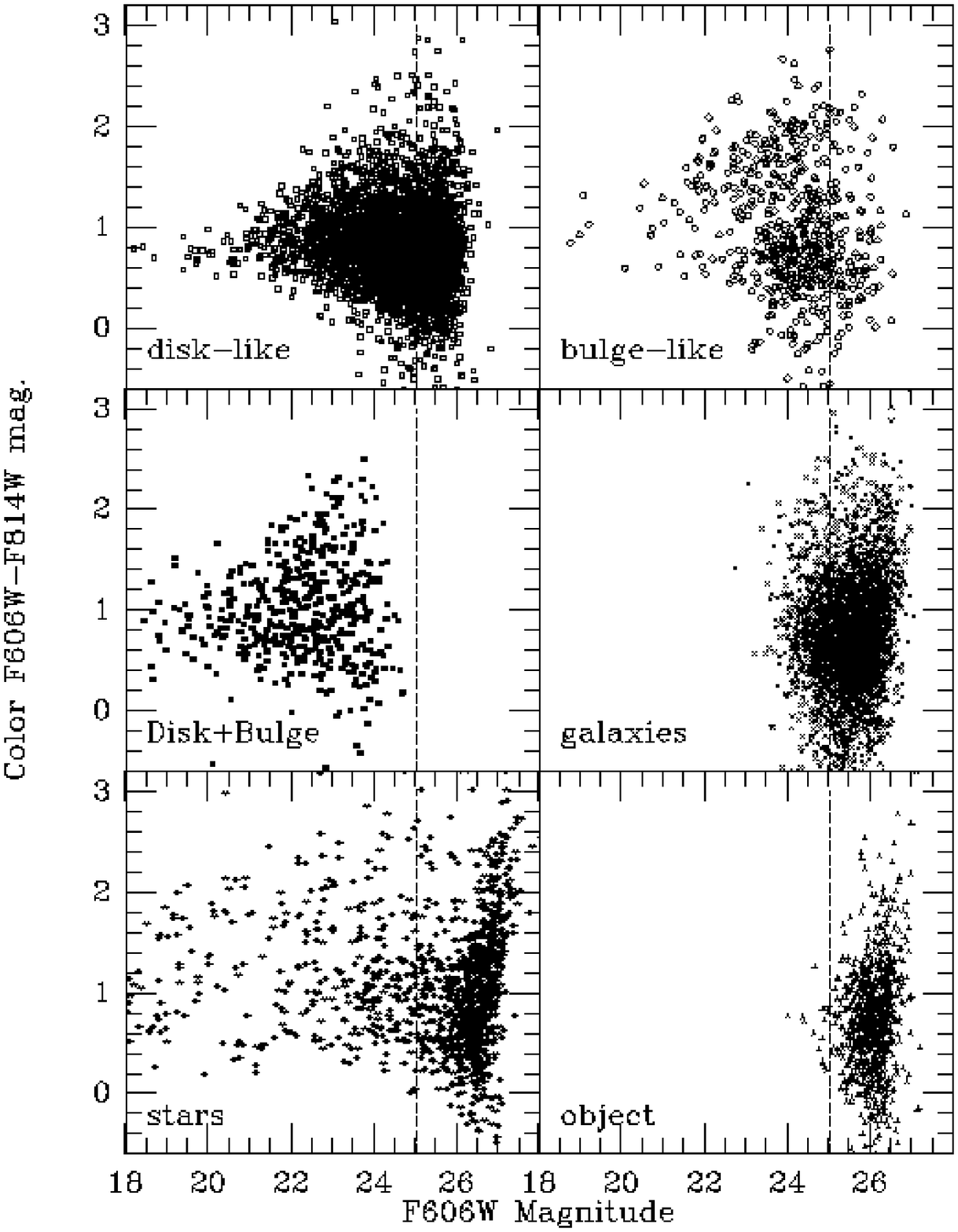}
\medskip
\caption{
Color of GWS galaxies as a function of the F606W apparent
magnitude. The dotted vertical line is drawn at the adopted the limiting
magnitude for morphology classification at $\SNRIL\approx 1.8$.
\label{fig22}}
\end{figure}

In Figure~\ref{fig22} we look at the color of GWS galaxies as a
function of the F606W apparent magnitude. We have shown all 6
classifications. The dotted vertical line is drawn at the observed
completeness magnitude ( $\SNRIL\approx 1.8$ ). Most of the objects which
were not classified morphologically are fainter than this
limit. Furthermore, some of the images fainter than this limit and
which have been classified as point-like are probably faint galaxies
rather than stars. We have chosen not to filter the MDS catalogs
brighter than this limit in order to avoid additional censorship of
the sample in statistical analyses. All parameter estimates for
galaxies fainter than this limit (i.e. magnitudes corresponding to
$\SNRIL \ltorder 2.2$) can be used for statistical analyses only, i.e. studies
should not be focused on individual galaxies, particularly any
outliers of such a distribution.

\begin{figure}
\epsscale{.75}
\plotone{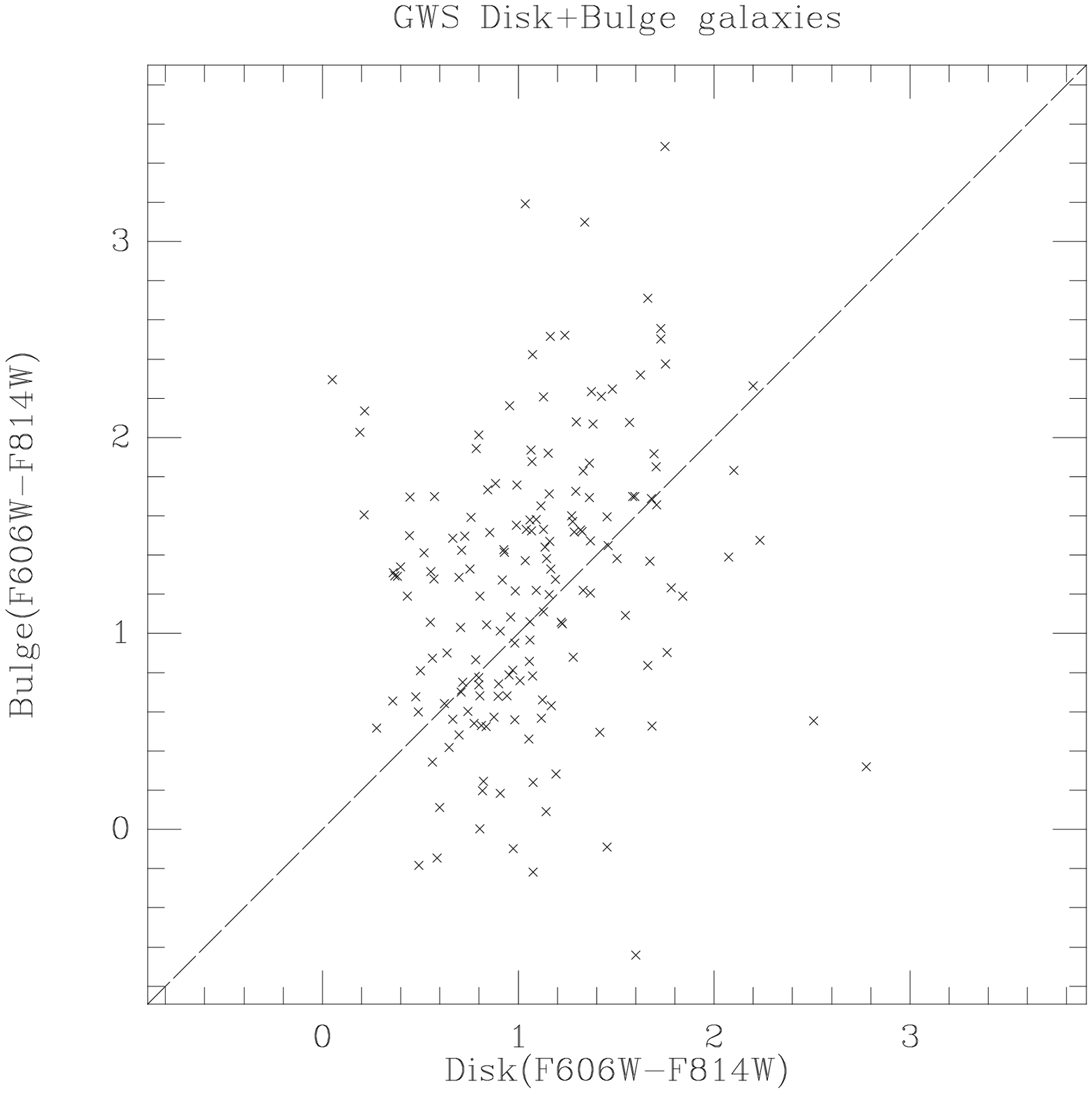}
\medskip
\caption{
Color of the bulge-components of galaxies with the corresponding
disk-components for galaxies in GWS which were fitted with \DPB
models in both F814W and F606W.
\label{fig23}}
\end{figure}

In Figure~\ref{fig23} we compare the color of the bulge-components of
galaxies with the corresponding disk-components for GWS galaxies which
were fitted with \DPB models in both F814W and F606W. It appears that
the colors of disk and bulge components of many galaxies are similar
(the dotted line), although bulges are observed to be systematically
redder, as expected, except for a few isolated cases.

\begin{figure}
\epsscale{.75}
\plotone{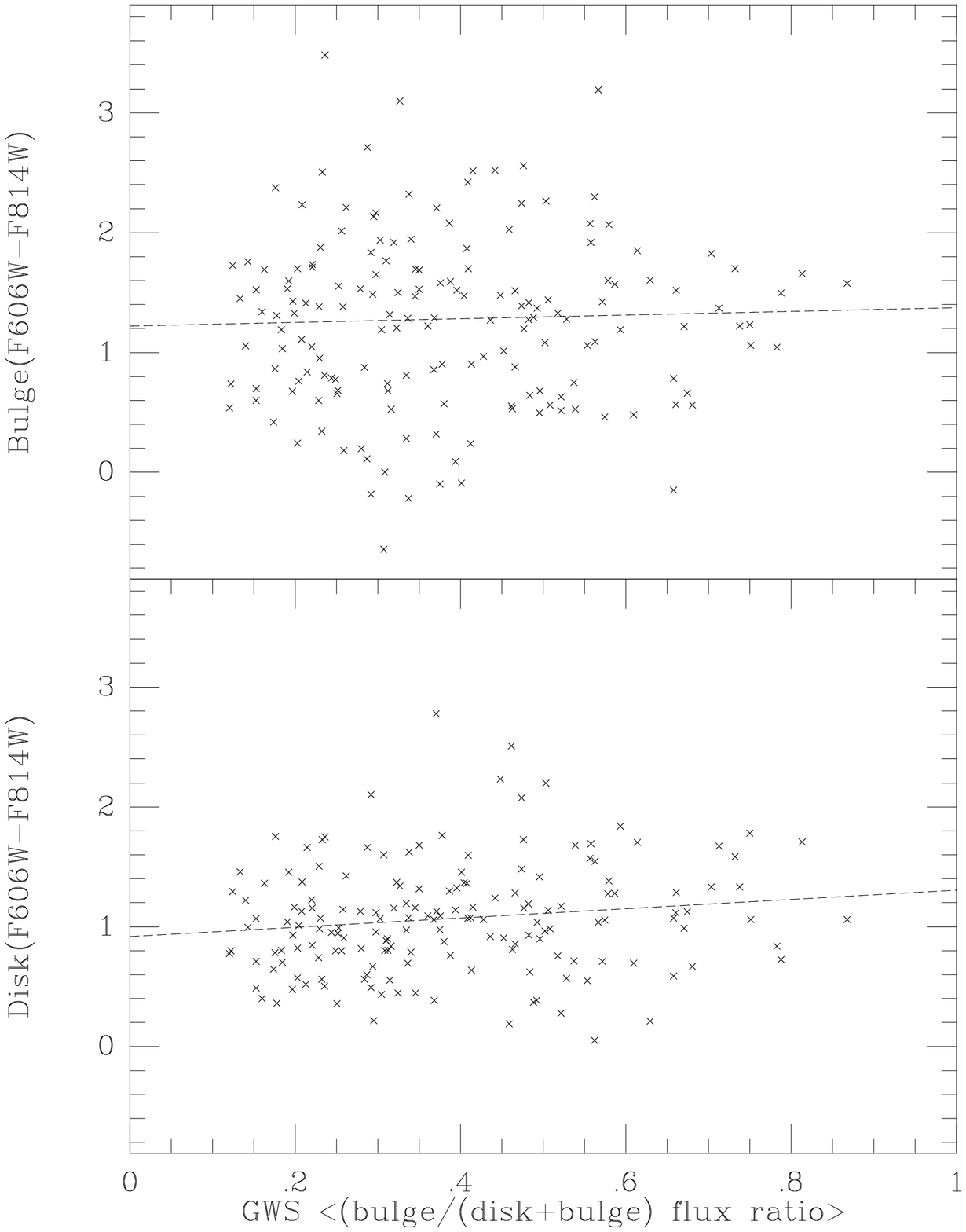}
\medskip
\caption{
The colors of the disk and bulge components as a function
of the $\BTD$ flux ratio for galaxies in GWS
\label{fig24}}
\end{figure}

In Figure~\ref{fig24} we look at the color of disk and bulge components of
galaxies as a function of the $\BTD$ flux ratio. As may be expected,
the disk components of galaxies appear to become $\approx 0.5$ mag
redder as we follow the plot from disk-like to bulge-like galaxies,
with a cosmic scatter of 0.45 mag. The colors of bulges remain
practically the same $\approx 0.2$ mag., with a larger cosmic scatter
of 0.6 magnitudes.

\begin{figure}
\epsscale{.75}
\plotone{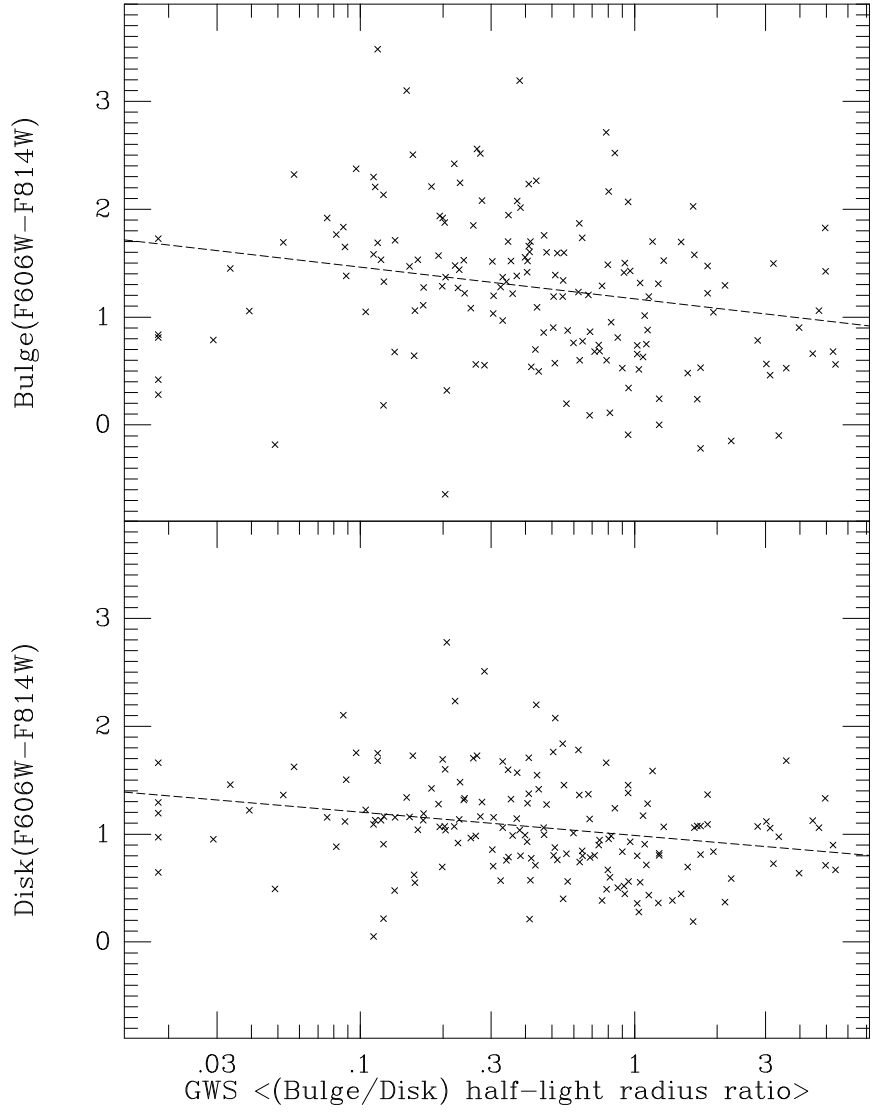}
\medskip
\caption{
The colors of galaxies in GWS as a function of the Bulge/Disk half-light radius
ratio ($\HLF$).
\label{fig25}}
\end{figure}

In Figure~\ref{fig25} we look at the color of galaxies as a function
of the $\HLF$ ratio. It appears that redder galaxies have a smaller
ratio. Careful statistical analysis is needed to ensure that this is
``real'' and is not caused by a selection effect in which the GWS
galaxies were sufficiently bright that \DPB models could be fitted to
images in both F814W and F606W filters.

\section{Conclusions}

An automated maximum Likelihood procedure has been developed to
calibrate, detect and quantitatively measure objects in the HST WFPC2
fields. The procedure measures the parameters of faint galaxies,
despite the potential difficulties related to the undersampling in
WFPC2.
 
\DPB models are now fitted routinely to the brighter galaxy
images as a part of the MDS pipeline. A \DPB galaxy model, a pure
disk, a pure bulge or a star model is chosen automatically using
likelihood ratio tests. Classification is done for images with
significant confidence.
 
Most HST MDS fields observed in 1994-1997 have been processed,
resulting in a catalog of over 200,000 objects which have been put on
the MDS website with a searchable browser interface. Clicking on a stack image
will pick out and display the maximum likelihood model fit and the
parameters for that object.
 
The statistical properties of the HST-MDS Catalog has resulted in many
publications and comparisons with models of galaxy evolution will
continue.
 
\acknowledgments
 
 This paper is based on observations with the NASA/ESA {\it Hubble
Space Telescope}, obtained at the Space Telescope Science Institute,
which is operated by the Association of Universities for Research in
Astronomy, Inc., under NASA contract NAS5-26555. The Medium Deep
Survey was funded by STScI grant G02684 {\it et seqq.} and by the HST
WFPC2 Science Team under JPL subcontract 960772, under NASA contract
NAS7-918. Some of the data was also processed under the STScI archival
grants GO6951, GO7536, and GO8384.

We acknowledge the multiple contributions of Dr. Stefano Casertano,
Dr. Myungshin Im, Mr. Adam Knudson and Dr. Lyman Neuschaefer who were
associated with the MDS pipeline processing and analysis. 
We also thank the rest of the original MDS Co-I team, including Dr. Richard
Ellis, Dr. Gerard Gilmore, Dr. John Huchra, Dr. Garth Illingworth,
Dr. David Koo, Dr. Antony Tyson and Dr. Rogier Windhorst for their
contributions to the program.

\centerline{\bf Appendix - MDS Pipeline}
\bigskip
\section{Association of WFPC data and MDS Field names}
 
The MDS database was maintained and updated using Starview 
(\cite{1994STScI.AM5...23F}).
Observations are assigned an alphanumeric 5-character name that is
based on Galactic coordinates as described below such that fields
which are from the same region of the sky are associated by name.
The choise of the individual characters in the name is as follows:

1) The first letter of the name, 'u', is the HST instrument letter
assigned to WFPC2 observations. It was 'w' for older WF/PC data.

2) Galactic Latitude from `a' in the south to `z' in the north in
equal steps of sin(latitude), using numeric index [6-9,0-5] within 16\degpt1
from the Galactic plane.

3) Galactic Longitude using sequence `1-5,a-z,6-9,0' in steps of 10\degpt0
such that numeric indexed fields are towards the Galactic Center. The
Galactic caps within 3\degpt9 from pole are assigned ``a-'' and ``z\_'' SGP
for NGP respectively.

4) Chronological sequence of primary target within the 31 
degree$^2$ cells defined above, based on coordinates. Observations within
a 0\degpt5 radius are assumed to be of the same target.

5) Chronological sequence of Association around the same primary
target set. These fields may overlap each other.

The program using a list of all pure parallel WFPC2 GO and GTO
observations assigns the names. We have not included the STScI
UV-survey program (pid=6253) or the current archive program (pid=7909)
for all parallel WFPC2 data since February 1998. Every dataset in an
associated group is allowed to be a maximum of 8\farcs0 (10\% of
WFC CCD width) from any other dataset in the same association. This
range is sufficient to associate all WFPC data taken in parallel with
a STIS or NICMOS observations, which are dithered, say within a 5\farcs6
square. For most cases the \PAV3 orientation is identical.
If it is not, then we ensure that the difference in rotation is less
than 0\degpt03 . This ensures a 1-1 mapping of the pixels, keeping
any effect caused by the small rotation or differential distortion to
be under about 0.5 pixels, the maximum error made by adopting integer
truncated pixel shifts between images in a stack.

\begin{figure}
\plotfiddle{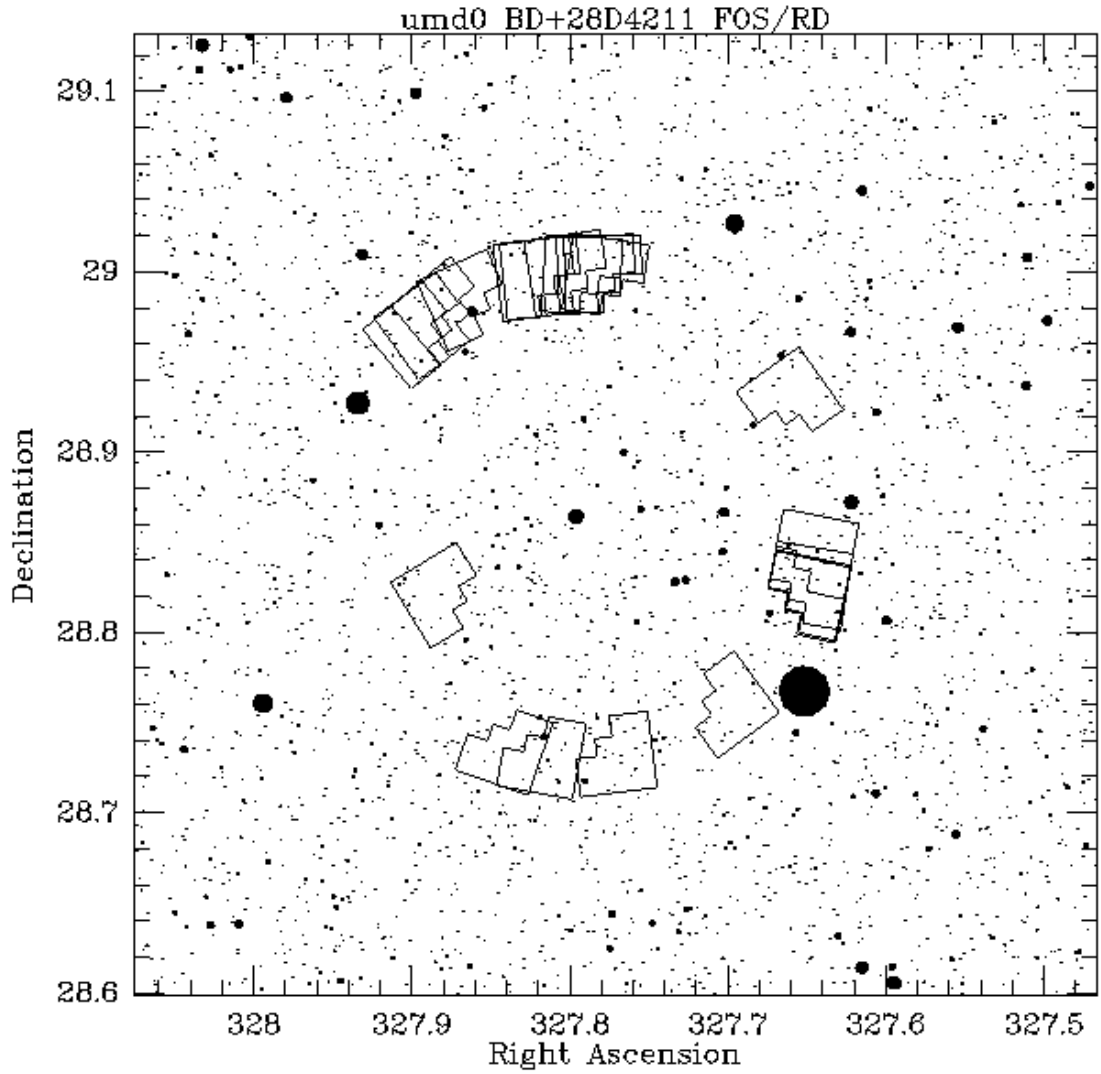}{6.0in}{-90.}{75.}{75.}{-288}{432}
\medskip
\caption{
Repeat observations give a ring of pure parallel WFPC2 fields around
FOS calibration star BD+28-4211.
\label{figA1}}
\end{figure}

Around some objects such as the FOS calibration star BD+28-4211 MDS
has many repeat observations as illustrated on Figure~\ref{figA1}.

\section{Calibration procedure}

Briefly, the calibration procedure is as follows.

The WFPC2 images are calibrated using the best available calibration
data. We adopt the STScI static mask, super-bias and super-dark and
flat field calibration files created for the HDF. Tables of hot
pixels from STScI are used to correct fluxes in fluctuating warm
pixels for the period of observation. Correction is made to ensure
that the noise from any residual warm current is smaller than the read
noise. Hot pixels, which cannot be corrected to that accuracy, are
rejected. Saturated pixels and pixels with large dark current are
flagged as bad and ignored. No attempt is made to interpolate over
them. The software has been specifically developed to recognize the
existence of missing pixel values.

In general we have more than one exposures in the same filter, in the
same field, to reject the numerous cosmic rays by stacking exposures
with a $3\sigma$ clip. We use a corrected version of the IRAF/STSDAS
combine task. See \cite{1994AJ....108.2362R} for a detailed
discussion of various aspects of the stacking procedure and the
statistical errors which are corrected in the ``combine'' algorithm.
A error image is also generated which computes the rms error from
the noise model, taking proper account of pixels rejected by cosmic
rays, the dark current, flat-field.

Shifts between images were determined by cross-correlation of the
images. The coordinates listed in HST WFPC2 image and/or jitter file
headers are often found to be insufficiently precise for the process
of image stacking (\cite{1997STScI.CT3..361R}). The shifts are
determined to an estimated rms accuracy of 0.1 WFC pixels. To avoid
interpolation (which spreads the charge from cosmic rays, charge that
is otherwise well confined), exposures are stacked with shifts
corresponding to the nearest integer number of pixels, without any
rotation or drizzling. Drizzled images (\cite{1997STScI.CT3..518F})
are most useful for very deep exposures like the HDF, which do not
occur in pure parallel observations. Drizzling causes the errors in
adjacent pixels in the image to become correlated and significantly
complicates a proper statistical analysis of the image.

A mode offset is employed to allow for changes in the sky background
in different exposures due to changes in the fluorescent glow and
scattered Sun/Earth light. The calibration accuracy is partly limited
by the fluorescent glow. This can contribute as much as 50\% of the
dark current, and is strongly correlated with the cosmic ray activity
during the WFPC2 exposure, which in turn depends on the particular
orbit. However, except for very deep stacks like the 
HDF, the noise created by improper correction of this
fluorescent glow results in a term which is small compared to other
noise terms.

We next remove any large-scale gradient from the faint outer regions
of bright galaxies, for which the nucleus was probably the target of
the primary observation. The four CCD images of the WFPC2 are first
oriented and merged along the pyramid edge and a single 2nd order
6-parameter polynomial surface is fitted across all four. This
surface is then subtracted from each of the individual images and this
automated procedure is iterated 2 or 3 times until no gradient is
visible. Only about 4\% of the processed MDS observations required
this gradient removal.

After stacking, the image is multiplied by a selected factor, which is
a power of 2, followed by an integer truncation and division by the
same factor. This makes the images compressible without any loss of
useful information, since the differential values before and after
this process are much smaller than either the accuracy of the
calibration or the averaged read-noise. The selected power depends on
NCOMB, the number of images stacked and we adopt the function ${\rm
nint}(\log_{2}(8\ {\rm NCOMB}))$, i.e. $2^3$ for a single image and
$2^6$ for a deep 6-stack. The estimated rms error has an expected
dynamic range of 0 to 25 ADU and the accuracy is unlikely to be better
than 0.01 ADU. Therefore the rms ADU error image is multiplied by 100
and truncated to the nearest integer, in order to generate a short
Integer image which is half the size when uncompressed than the size
of the corresponding image of real numbers.

\section{Object detection}

The MDS pipeline was used to process only the typical field
in which crowding was not a problem. We selected sparse fields in
which the number of pixels $4\sigma$ or more above sky is typically
under 5\% of the total pixels in the field. We classified as non-survey
and excluded from the MDS pipeline image analysis all low galactic
latitude fields with lots of stars, and those fields close to Globular
Clusters and local Group galaxies with, say, more than about 1600
objects detected. Non-Survey MDS fields were analyzed independently
by other members of the team.

\begin{figure}
\epsscale{.75}
\plotone{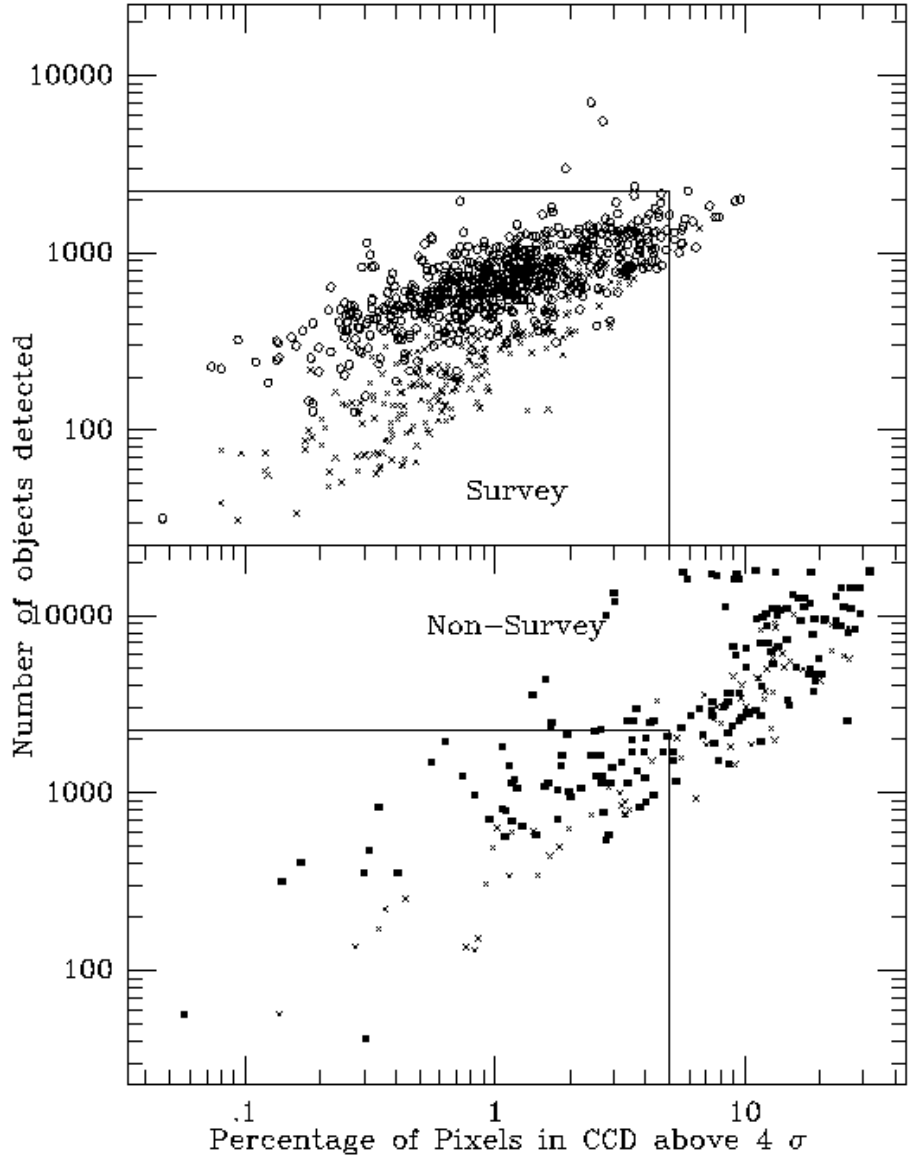}
\medskip
\caption{
Number of objects detected as a function of of pixels above $4\sigma$.
The fields selected to be part of MDS survey have less than about 5\%
of pixels above $4\sigma$. WFPC2 images with no cosmic-ray split have
been indicated with crosses to show the systematic smaller number of
object detections because of the attempted cosmic ray cleaning.
\label{figA2}}
\end{figure}

In Figure~\ref{figA2} we illustrate the number of objects detected as a function
of the number of pixels above $4\sigma$. The figure shows that most of
the fields selected as part of the MDS survey have a smaller fraction
of pixels over $4\sigma$ and a smaller number of objects detected than
in non-survey fields. In both cases, images with no cosmic-ray split
have been indicated with crosses and show a systematic smaller number
of object detections because of the attempted cosmic ray clean out.

Objects are located independently on each image using a `find'
algorithm developed for HST-WFPC data. This algorithm does not do any
pre-convolution of the data, so that it is specifically designed to be
insensitive to hot pixels and missing pixel values. It is based on
finding local maxima and mapping nearby pixels to the central object,
and then selecting those detections which are significantly above
noise. The detection threshold algorithm originally developed for
pre-refurbishment WF/PC data was optimized for WFPC2. This resulted
in the location of a practically identical list of objects in the
overlapping region of three WFPC2 MDS parallel fields USA0[1-3]
observed in June 1994. To ensure that we do not break up bright
galaxies into small regions of star formation, we adopted an object
resolution of $0\farcs5$, and small regions within this radius
were allowed to merge with a brighter center. A larger radius of $1\farcs0$
was used for WF/PC data. This algorithm has been
observed to locate real objects with as much or better efficiency as
the FOCAS algorithm (\cite{1979ApJ...230L.153T}), which was developed mainly for
ground-based data.

The MDS `find' algorithm generates both a catalog and a `mask' image,
which associates each pixel with one object. This is a short integer
file since the MDS pipeline assumes that there are less than 10K
objects in a single WFPC2 field suitable for analysis. The stacking
and initial object location procedure is a fully automated first step
of the MDS pipeline. After the initial find, we have the only
interactive part of the operation. We first look at the exposure, and
confirm that it satisfies our requirements on inclusion into the
MDS. A typical MDS survey field is uncrowded, with
about 400-800 objects detected in the 5 arc-min$^2$ field. We also
exclude from the MDS catalogs those objects with a centroid within 10
pixels ( $1\farcs0$ ) from the pyramid and CCD edge, thus
reducing the area surveyed by about 5\% from 5.03 arc-min$^2$ to
4.77 arc-min$^2$ per WFPC2 field, and causing a $2\farcs0$ wide gap
in the shape of a cross in the center of each field. Rapid changes in
the image distortion and the PSF ( a residual consequence of the
original HST Spherical Aberration ) make the edge a very difficult
region for reliable quantitative analysis.

The next operation is to fix up the mask for any bright objects which
have been over resolved, or to delete any ghost images or extremities
of bright stellar diffraction spikes which have been spuriously
detected as objects. The detection algorithm has been optimized to
work best at intermediate to faint magnitudes at the cost of over
resolving a few bright objects. The numbers plotted in Figure~\ref{figA2} are raw
counts before cleanup. The spurious detections are flagged with an
interactive cursor for rejection or merger with the central image.
This interactive operation takes about 30 min per stack and is done
with a well defined set of guidelines which were originally developed
for WF/PC data and modified appropriately for WFPC2.

The object detections in the various filters are then matched by
software, and a single catalog is created, together with a revised
mask for each image, so that corresponding pixels in the different
filters are associated with the same object. Looking at a grid of the
individual object detections, the final masks are inspected and the
procedure is iterated as required to ensure that the object
definitions as encoded by the final masks are acceptable. This is the
conclusion of the calibration and object detection phase of the MDS
pipeline.

The object detection algorithm and search thresholds were kept
unchanged over the four years of the MDS. When new calibration data
became available we recalibrated the data and created a new stack to
obtain a slightly lower noise in the image. We however do not
redetect objects, however. The masks remain constant, and after the
field has been setup in the MDS database, model fits to any objects
can be reprocessed with practically no human intervention. When there
has been a significant improvement in the calibration or the fitting
software, the whole database is reprocessed to obtain an improved
version of the catalog, which is uniform over the whole period of
observation. This has, however, become practical only after we
obtained a SPARC Ultra-1, which on its own can do the reprocessing of
the current database of over 400 fields in about two months. All of
the MDS fields have been reprocessed with the last (July 1996) version
of the MDS image analysis software. The shifted stacks were refitted
after they were improved in July 1997 using inter-image shifts derived
from cross-correlation analysis.

\section{Definition of the object region for analysis}

Most galaxies are analyzed by picking out a 64-pixel square region
centered on the galaxy. The very few images (on average about 3
galaxies per WFPC2 field or 0.67\% of the catalog) which are larger
are analyzed as 128-pixel square images and in an extreme case as
256-pixel square region. The integral power of two in the region size
in pixels was chosen for efficient convolution of models by fast
Fourier transforms (FFT).
 
An initial guess of the local sky background is determined from an
algorithm that determines the sky using an adaptation of the
iterative, asymmetric clipping procedure as described by
\cite{1984AJ.....89..176R}. In the very few cases for which it is
detected that the local sky is poorly defined (large rms and skew in
distribution) then the global sky is adopted as the initial guess. We
next use the mask of detected objects generated by the MDS `find'
program in order to define a $1\sigma$ contour around the object. This
is done by selecting the subset of pixels which are next to each other
and are $1\sigma$ above the estimated local sky.

Despite careful `dark' calibration and correction for suspected hot
and warm pixels, pixels with fluctuating dark current are seen to
leave a few ``hot'' pixels in the image. Since they could contribute
significant flux compared with the flux of some of the faint images,
any isolated pixels in the region outside the $1\sigma$ contour and over
$5\sigma$ above immediate neighbors were located and were assumed to be
hot pixels and rejected. This algorithm detected hot pixels in
only 25\% of the images and in these cases found on average only
5-pixels in a 64 pixel square region (See Figure~\ref{fig7}). These values are
for the GWS taken with the WFPC2 before it was cooled down from $-78^o
C$ in April 1994, and for which warm-pixel corrections are not
available. There are many less hot pixels in the newer data taken at
$-88^o C$.

The initial guess of the local sky and the choice of pixels within the
$1\sigma$ contour associated with the object are factors which influence
only the region picked out for analysis. The pixels within the
$1\sigma$ contour get no different treatment when the likelihood
function is integrated.

\section{The observational error distribution.}

The presence of cosmic rays makes the observational error distribution
of the raw observation non-Gaussian. We have found that the cosmic
ray contamination can be represented by a Weibull distribution with
index 0.25. In theory, the likelihood function can be defined by
taking the model all the way back through the calibration procedure in
order to make the comparison by summing over independent raw
observations without any stacking. If this is done, one can take
proper account of the effect of telescope ``breathing'' which results
in slight changes to the observed PSF. One can also allow for
contamination by faint cosmic rays and even any analog to digital
conversion errors (a problem mainly for old WF/PC data) on the
observational error distribution (\cite{1994AJ....108.2362R}).

However, after extensive software development and investigation using
simulations, we found that this analysis of raw observations and the
use of a complex error distribution gained only about 0.15 magnitudes
in quality of morphology classification over the very much simpler
analysis of calibrating and stacking the image to remove cosmic rays
and the assumption of a Gaussian error distribution. With the latter
approximation the log likelihood function is equal to
$$-0.5(1+\log(2\pi))\chi^2=-1.42\chi^2 .$$ Maximizing the likelihood
function is then identical to minimization of $\chi^2$. We have
currently chosen to use the simpler analysis since the very slight
improvement in results does not justify the very large increase in
computation.

\section{Generation of model images for comparison with observation.}

The creation of the model image is the most technical and computer
intensive part of the procedure. On average, of order 700 model
images are used by the minimization routine to converge on the
best-fit model of a single object. Since our minimization routine
uses derivatives, an efficient high precision algorithm is required.
For under-sampled images like those from WFPC2, sub-pixelation is very
important, particularly close to the central peak of the galaxy image.
We have developed a procedure which is automatically optimized by the
algorithm by testing the evaluated likelihood function on the image
being analyzed. We find that for many images the central pixels of
the model image convolution needs to be done in sub-pixel space, and
then block averaged for comparison with observation.

\section{The creation of the image.}

In order to ensure that the evaluated likelihood is a smooth function
of all the model parameters, require computation of the model image at
much higher resolution than that observed, particularly in
under-sampled regions close to the center. The image models we have
adopted are scale free and have an axis of symmetry. In order to
minimize computation and make use of this symmetry, we therefore first
evaluate the image by adopting an origin at the middle of central
pixel of the array. The outer regions of the image are evaluated
without sub-pixelation. If the models are scale free, then the outer
regions can be multiplied by a constant factor to obtain sub-pixeled
values for the inner pixels.

For example, we generate an 81 pixel square image by first computing
pixels outside the inner 27 pixel square. Using the axis of symmetry,
only half these pixels need to be computed. Then each pixel outside
the inner 9 pixel square and within the 27 pixel square region is
integrated with 3x3 subpixelation by integrating the 9 pixels at 3
times the radius and using a scale factor appropriate for the selected
model. Following this step, each pixel outside the inner 3 pixel
square and within the 9 pixel square can be integrated with an
effective 9x9 subpixelation, and the region outside the central pixel
and within the inner 3 pixel square can be integrated with an
effective 27x27 subpixelation. Finally the central pixel with an
81x81 subpixelation is integrated from the rest of the whole image and
the contribution for the very central subpixel. In this way the model
image that is created has a very high degree of subpixelation for the
inner pixels at practically no extra computation. In this example,
the central 64-pixel square region used gets computed as a 280 pixel
square image with increasing subpixelation towards the center. The
image is effectively computed at 39200 points at the cost of 2917
evaluations, more than an order of magnitude increase in speed. This
approach of sub-pixelation can be used on any scale free model, even
if not elliptical.

\section{The Point Spread Function}

Selection of the Point Spread Function (PSF) is not easy. The choice
is between, using observed PSF's of well-exposed stars, or model PSF's
from programs such as tinytim (\cite{1992STScI.TTM....1K}). 
Observed stellar PSF's are 
under-sampled, have random observational errors, and are not always
available close to the image being analyzed. A compiled grid of
stellar PSF's from various observations has systematic errors
comparable to the small systematic errors seen in model PSF's.
tinytim PSF images have the added advantage of being able to be
generated as a sub-sampled image without observational jitter or the
scattering in the WFPC2 CCD photon detection (see below for details).

Convolution of the WFPC2 model image is best done in sub-pixel space
where it is less under-sampled. Tinytim (\cite{1992STScI.TTM....1K}) 
PSF's are evaluated with 3
and 5 times sub-sampling for the PC and WFC CCD chips respectively.
The 267 square PSF images are stored in the same data file format as
the observations in a 3 by 3 PSF grid for each chip, and centered on
the image at the pixel for which they were evaluated. A PSF grid
image data file is made for each filter used in the observations. In
the image analysis we choose from the grid the PSF for which the
center is nearest to the location of the object. A 3 by 3 grid is
sufficient for the corrected optics of WFPC2. A 11 by 11 grid with no
sub-sampling was used for pre-refurbishment WF/PC data for which
under-sampling of the extended PSF due to the spherical aberration was
relatively less of a problem than the rapidly changing PSF as a
function of the location on the chip.

We have so far ignored the changes to the PSF caused by the gradual
shift of the mean focus between resets (maximum of 6 microns) and the
telescope breathing which has a rms of ( 3 microns ). This in itself
is a complicated issue when a stacked image is used in the analysis.
The focus of every exposure in the stack cannot be assumed to be the
same. Simulations have shown that for typical extended images with a
half-light radius larger than say 2-pixels, model parameters derived
using slight changes to the PSF are well within the parameter error
estimates. For extended images, deviation of a galaxy from the simple
model assumed could give larger errors to the parameter estimates.

\section{Convolution}

Convolution is done with IMSL 9.2 FFT routines. For most images a 64
pixel square array is used. We correct for any aliasing by generating
a 16 pixel square image with a factor of 4 lower resolution and
convolving it at the center of a 32 pixel square image where the
region outside the central 16 pixel square is set equal to zero. The
flux which gets convolved out to this border region is a sufficient
estimate of the alias, and subtracted appropriately from the
convolution of the original 64 pixel square array which is not
surrounded by a zero border to prevent aliasing. This correction
procedure takes only 1.25 times longer, rather than the 4 times
increase needed to surround the 64 pixel square array with a zero
buffer, and doing a 128 pixel square array convolution.

The subpixelation chosen is such that the region of the image selected
for analysis will fit within a 64-pixel square array at the subpixel
resolution. However, for WFPC2 images this was found to be
insufficient for many highly peaked images that also covered most of
the 64-pixel square array without any subpixelation. In these cases,
we generate a second image for just the central pixels at the
sub-pixelation used for PSF and then convolve this image as a 32-pixel
square array, and correct it for aliasing effects using a similar
procedure as for the main image.

The high-resolution image replaces the central 5-pixel square region
of the model image. Since the center region after convolution is at
5-pixel subpixelation, the image can be shifted to the required center
and block-averaged to the observed pixel scale using a simple
algorithm that assumes a uniform flux distribution within each
subpixel. Including such a higher-resolution center takes factor of
1.25 longer in CPU time, and used as required based on changes to the
likelihood function.

\section{Scattering at time of photon detection in WFPC2 CCD.}

There is a non-negligible probability that a photon will be counted in
a pixel adjacent to that in which it should have been detected. For a highly
peaked source such as a stellar image, the location of the centroid
within a pixel will govern the spill over to the adjacent pixels. The
photon detection scattering of the model image therefore needs to be
done at the observed image resolution and cannot be incorporated
within a sub-sampled PSF.

After shifting the center and block averaging down to the size of the
observed pixels, the image is convolved by a 3-pixel square kernel to
allow for the photon detection scattering in WFPC2 data. The
symmetric kernel adopted has [center,side,corner] values of
[0.75,0.05,0.0125] and is as recommended in tinytim 4.1. We have
compared this kernel with an azimuthally averaged kernel like that
which was recommended in tinytim 4.0 with elements
[0.5628,0.0937,0.0156]. We find that the revised, more centrally
peaked kernel yields better model fits to a sample of stars. However,
the combination of PSF and scattering is still not perfect and leaves
residuals of about 5\% to 10\% which are significant in the brighter
images.

\section{Image Jitter}

Since the tinytim PSF models are generated without including any
contribution from telescope jitter, the intrinsic half-light radius
estimated for a point source is non-zero. For a WFPC2 primary
pointing in fine lock we expect about 10 mas (milli-arc-seconds) of
telescope jitter. However, for parallel WFPC2 observations it could be
larger because aberration corrections made for the primary instrument
are slightly different from those required for the WFPC2. Since jitter
data is still not available for all WFPC2 observations, they have 
as yet not been incorporated them into the MDS pipeline ( see
\cite{1995STScI.CT2..351R} for details).

Shifts between images were determined by cross-correlation of the images.
Any small sub-integer shifts between the images that are ignored in
the stacking procedure would also increase the effective jitter in the
stacked image. Pointing Errors of 10 mas are possible at the times of
target reacquisition between consecutive orbits of the HST and 20 mas if
the target is reacquired after some other observation.

Any systematic radial errors between the actual PSF for the
observation and the tinytim PSF model would also translate to a larger
effective half-light radius. We typically estimate about 20 mas for
unsaturated stellar images, growing to as large as 60 mas for the very
bright saturated stars. The half-light radius computed by the program
is not corrected as yet for jitter this correction which not have any
significant effect, except on those images with a half-light radius
smaller than a pixel.

\section{The flow chart of the fitting and the output files}

 The input data for the program are four STSDAS/GEIS images for each
filter, the calibrated image, corresponding error image, object
definition mask image, the PSF grid image, and an ASCII data file with
keyword information about the pointing and global noise
characteristics of each calibrated stack. These files are identified
in the header of the catalog that then identifies the object to be
analyzed by the group and coordinate of the centroid, together with
the mask number. In a special mode, it is possible if necessary to
identify a small group of adjacent objects resolved with different
mask numbers as a single object in the analysis.

 The program fits all available images of the object in the different
filters and outputs an ASCII data file with the fitted parameters,
covariance matrix and other information about the likelihood ratio, and
the sequence of intermediate results and tests. Catalogs for all
objects in a field or a number of fields can be obtained using a
keyword search of these data files. Also created is a FITS data image
for each object. This file has the format of a grid with a single row
of 7 images for each filter, starting from the longest wavelength at
the top and progressively shorter wavelengths below it.

 From left to right the images on a single row are

  (1) Full image area read from the stack as observed.

  (2) Selected region for analysis, with any adjacent objects masked out.

  (3) Maximum likelihood model image, following PSF convolution.

  (4) Maximum likelihood model image.

  (5) Residual image.

  (6) Error image.

  (7) Object mask image.

To make the FITS files short integer and compressible, the sky
subtracted stack images and the residuals are multiplied by 10 and
transformed to short integer. The error image and mask are in the
integer format used for these images (see sec~15).

\begin{figure}
\epsscale{.75}
\plotone{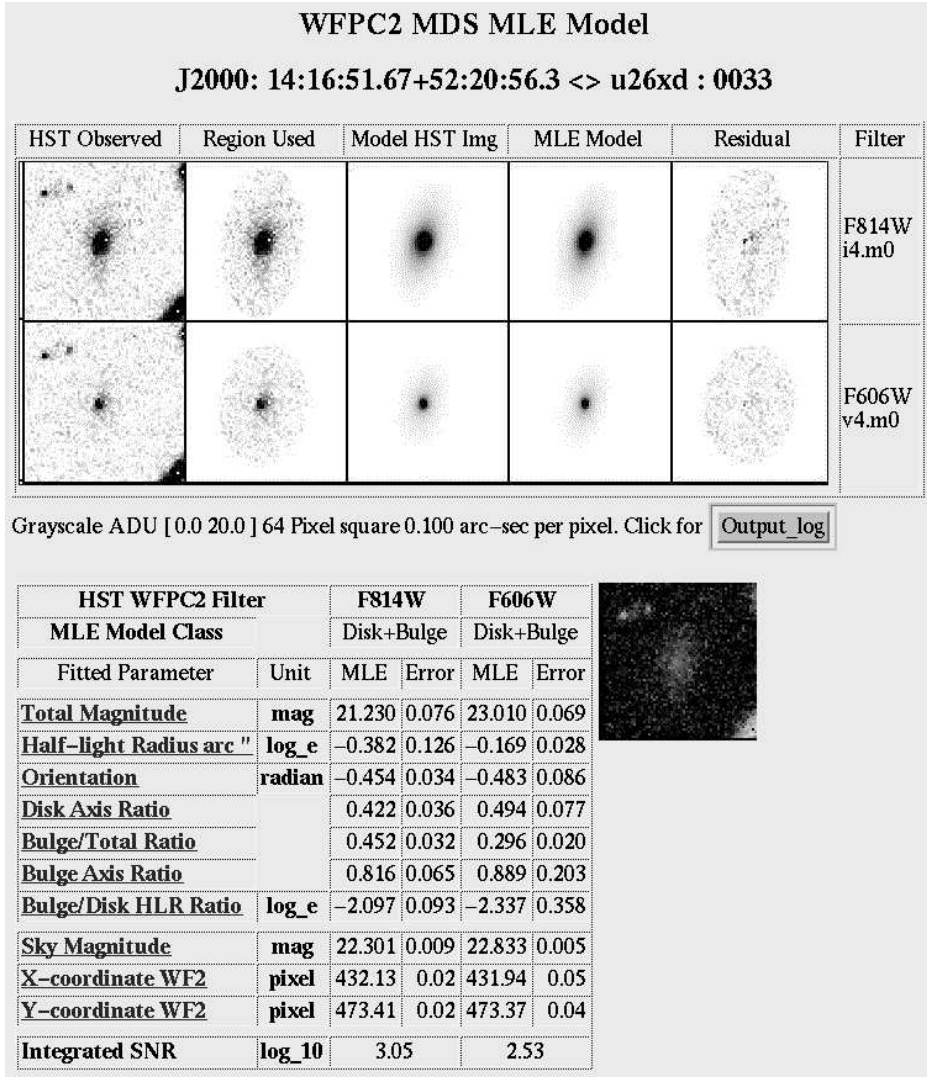}
\medskip
\caption{
An example of a well exposed galaxy image from the MDS database as
displayed via the MDS website. The dark image is in color and is
generated from the available filters using fixed color transformation
algorithm.
\label{figA3}}
\end{figure}

We show on Figure~\ref{figA3} an example of a well exposed galaxy
image from the MDS database as displayed on the MDS website. Note that
even the fainter F606W image which has $\SNRIL\approx 2.5$ is in the
range signal-to-noise index which gives reasonable \DPB decomposition.

As of October 1998 similar output for over 200,000 galaxies and stars
had been made available on 19 CDROMS installed on a `Jukebox' at STScI.

Analysis of an image as a star, disk-like or bulge-like galaxy is
fairly straightforward. The only parameter that may be dropped is the
axis ratio and in that case the orientation parameter is also not
needed and the number of parameters fitted drops from 7 to 5.

The software can also use any profile index or even attempt to
optimize its value as was done for images in the Uppsala galaxy
catalog (\cite{1989ESO...XXX....1L}). However numerical simulations
have shown that the profile index gives a measure of the $\BTD$ flux
ratio only if the axis ratio of the two components are very
similar. Else, the minimization procedure computes an index which is
not within the range of one (for pure disk-like) and a quarter for
(pure bulge-like). The value is often larger than two, as seen for the
Uppsala galaxy catalog in which the index seems to have been
constrained to be smaller than three for the same reason.

The \DPB analysis is much more complicated. From the very choice and
definition of the fitted parameters, to the automated selection of
them to ensure convergence, has been a long investigation based on
both fits to real data and to realistic simulations. Almost like the
minimization process, getting close to the answer was much faster than
checking and ensuring that the algorithm was optimum. Getting an
algorithm that worked on a majority of the images has been tested and
in use for sometime. The final optimization became practical with the
aid of a SPARC Ultra-1 which is more than an order of magnitude faster
than a SPARC-2 on which the programs were developed.

Automation gives uniformity at the cost of a few complicated cases
(particularly at bright magnitudes) where a decision made by human eye
would probably be different. The program was improved constantly till
about July 1996 to reduce the percentage (currently about 2\% ) of fits
which are in error. All of the MDS database has been reprocessed with
the improved version of the program logic.

\end{document}